\documentstyle[onecolumn]{mn}
\newcommand{\mincir}{\raise -2.truept\hbox{\rlap{\hbox{$\sim$}}\raise5.truept
\hbox{$<$}\ }}
\newcommand{\magcir}{\raise -2.truept\hbox{\rlap{\hbox{$\sim$}}\raise5.truept
\hbox{$>$}\ }}
\newcommand{\minmag}{\raise-2.truept\hbox{\rlap{\hbox{$<$}}\raise 6.truept\hbox
{$>$}\ }}
\newcommand{\be}{\begin{equation}}
\newcommand{\ee}{\end{equation}}
\newcommand{\ba}{\begin{eqnarray}}
\newcommand{\ea}{\end{eqnarray}}
\newcommand{\brr}{\begin{array}}
\newcommand{\err}{\end{array}}
\newcommand{\bc}{\begin{center}}
\newcommand{\ec}{\end{center}}
\newcommand{\br}{\mbox{\bf r}}
\newcommand{\bv}{\mbox{\bf v}}

\newcommand{\bq}{\mbox{\bf q}}
\newcommand{\bx}{\mbox{\bf x}}

\newcommand{\hm}{\,h^{-1}{\rm Mpc}}

\title[Topology of the Cluster Distribution]{The Cluster Distribution
as a Test of Dark Matter Models. IV: Topology and Geometry}
\author[R. C. Pearson et al.]{P. Coles$^{1}$, R. C. Pearson$^1$,  
S. Borgani$^{2,3}$, 
M. Plionis$^{4,3}$ \& L. Moscardini$^5$\\
$^1$ Astronomy Unit, School of Mathematical Sciences,
Queen Mary \& Westfield College, Mile End Road, London E1 4NS\\
$^2$ INFN Sezione di Perugia, c/o Dipartimento di Fisica dell'Universit\`{a},
via A. Pascoli, I--06100 Perugia, Italy \\
$^3$ SISSA -- International School for Advanced Studies,
via Beirut 2--4, I--34013 Trieste, Italy \\
$^4$ National Observatory of Athens, Lofos Nimfon, Thesio, 18110 Athens, 
Greece  \\
$^5$ Dipartimento di Astronomia, Universit\`a di Padova,
vicolo dell'Osservatorio 5, I--35122 Padova, Italy}

\begin{document}

\maketitle

\begin{abstract}
We study the geometry and topology of the large--scale
structure traced by galaxy clusters in
numerical simulations of a box of side 320 $h^{-1}$ Mpc, and compare them with
available data on real clusters. The simulations we use are generated by 
the Zel'dovich approximation, using the same methods as we have used
in the first three papers in this series. We consider
the following models to see if there are measurable
differences in the topology and geometry of the superclustering they
produce: (i) the standard CDM model (SCDM); (ii) a CDM model with
$\Omega_0=0.2$ (OCDM); (iii) a CDM model with a `tilted' power
spectrum having $n=0.7$ (TCDM); (iv) a CDM model with a very low
Hubble constant, $h=0.3$ (LOWH); (v) a model with mixed CDM and
HDM (CHDM); (vi) a 
flat low--density
CDM model with $\Omega_0=0.2$ and a non-zero
cosmological $\Lambda$ term ($\Lambda$CDM). 
We analyse these models using a variety of statistical tests
based on the analysis of: (i)
the Euler--Poincar\'{e} characteristic; (ii) 
percolation properties; (iii) the Minimal Spanning Tree construction.
Taking all these tests together we find that 
the best fitting model is $\Lambda$CDM and, indeed, the others 
do not appear to be consistent with the data. 
Our results demonstrate that despite
their biased and extremely sparse sampling of the cosmological density 
field, it is possible to use clusters to probe 
subtle statistical diagnostics
of models which go far beyond the low-order correlation functions
usually applied to study superclustering.
\end{abstract}


\section{Introduction}
The study of the distribution of matter on the
largest scales amenable to observation can provide important 
constraints on models of the formation of cosmological structures.
In particular, it has now become well established that a very
accurate and efficient way of describing very large scale structure
in the galaxy distribution is obtained by not looking at galaxies
themselves but at rich clusters of galaxies.
If the `standard' model of structure formation --  the
gravitational instability picture -- is correct, 
the expected displacements of  galaxy clusters 
 from their primordial positions are much smaller than
the typical separation of these objects. In principle, therefore, clusters
of galaxies can yield clues about the primordial spectrum of perturbations
that gave rise to them, without such clues being trampled on by the
effects of non--linear evolution.  Moreover, because clusters represent
highly overdense regions in the cosmological density field, these
objects display an enhanced clustering signal relative to that of
galaxies on the same scale, an effect usually known as biasing 
(Kaiser 1984).

This is the reason why so much effort has been
devoted to compiling deep cluster surveys, starting with the 
pioneering work of Abell (1958), Zwicky et al. (1968)  and Abell, Corwin \&
Olowin (1989), and leading up to  extended redshift surveys both in the
optical (e.g. Postman, Huchra \& Geller 1992; Dalton et al. 1994; Collins
et al. 1994, and references therein) and in the X--ray (e.g. Nichol, Briel
\& Henry 1994; Romer et al. 1994; Ebeling et al. 1996) regions of the spectrum. 

The properties of galaxy clusters may help to resolve some of the
issues that have led to the present relative stagnation in the
theory of structure formation. Since the demise of the standard
model of the 1980s -- the standard Cold Dark Matter model (SCDM) --
a number of contending theories have been proposed which are in better
agreement with the observations than SCDM but between which it is
difficult to discriminate using present observations of galaxy
clustering and the cosmic microwave background; for a review,
see Coles (1996). It is therefore important to try to find statistical
diagnostics of clustering that may reveal differences between these
models and the data to see if they do indeed explain the
details of the observed clustering phenomenon, 
as well as between the models themselves so one
can understand how the various extra ingredients involved in these
models alter specific characteristics of the clustering pattern.

Simple two--point statistical descriptions of superclustering
(i.e. the clustering of galaxy clusters)
have already yielded important clues about the shape of the
matter power spectrum on large scales (e.g. Peacock \& Dodds 1994;
Borgani et al. 1997)
and, more recently, this has been extended to simple properties
of the higher--order moments (e.g. Plionis \& Valdarnini 1995;
Plionis et al. 1995; Borgani et al.
1995; Gaztanaga, Croft \& Dalton 1995). However,
the complete statistical characterisation of the clustering
requires knowledge of all the higher order moments or,
equivalently, knowledge of the complete set of $n$--point correlation
functions (Peebles 1980). Such a description is extremely laborious to 
construct, tends to be swamped by discreteness effects and sampling
errors even at quite small $n$ and is in any case rather difficult
to interpret geometrically.

For these reasons it is useful to seek a description of clustering
which by-passes this more orthodox approach and looks for intrinsically
geometrical or topological signatures. One can hope that such approaches
might lead to robust quantitative 
descriptions of the void-filament network which is visually
apparent in the distribution of galaxies, and to relate this visual
appearance to the interaction of non--linear gravitational dynamics
on an initial density field with some assumed power spectrum. The hope
is therefore to pick out differences between models which are hard
to discern in measures such as the power spectrum.
Various approaches to this question have been
suggested and some of
them have been more successful than others in their application
to the data. One particular problem such descriptors face when they
are applied to superclustering, for example, is that these objects
are extremely rare and there are strong shot-noise effects which have
to be compensated for in some way.

In this paper, we aim to investigate a particular set of
topological or geometrical descriptors of the pattern
present in simulated cluster distributions and, where possible,
to compare the results from simulations with the analogous
results from the Abell/ACO cluster catalogue. We should stress
at the outset that this is an exploratory work and there are
reasons to suspect that the task of discriminating between
these models and the data might be extremely difficult.
First there is the problem of shot-noise we alluded to above.
Secondly, the available cluster sample is quite small and may
suffer from unknown selection effects. One can hope, however, that
better controlled cluster samples may emerge fairly soon from
ongoing galaxy redshift surveys. Third, it is extremely difficult
to construct sufficiently large $N$-body simulations of galaxy clustering
and select the appropriate clusters within them in the same
way that clusters are selected observationally (e.g. Bahcall \& Cen 1992;
Croft \& Efstathiou 1994; Eke et al. 1996). And finally, there is
the ubiquitous problem of understanding how the objects one sees
relate to the distribution of matter one calculates, a difficulty
generically known by the name of biasing and which was first discussed
in the context of rich clusters by Kaiser (1984).

In the spirit of exploration, therefore, we shall use simplified
models of superclustering, generated by using a method based
on the Zel'dovich approximation. This method has been used in a number
of previous studies of the distribution of clusters in both
position and velocity space (Borgani, Coles \& Moscardini 1994;
Plionis et al. 1995; Borgani et al. 1995; Tini Brunozzi et al.
1995; Moscardini et al. 1996; Borgani et al. 1997) 
and is known to be accurate in
comparison with the full $N$--body approach, provided the degree
of non--linear evolution 
at the scale of individual clusters
is not too strong. 

The outline of the paper is as follows. In Sections 2 and 3 we briefly 
describe our simulation method and the observed Abell/ACO cluster
sample, respectively. We then go on to discuss the various
clustering descriptors we use to analyse these data sets. First,
in Section 4, we discuss the topological properties of the 
isodensity regions in the distribution traced by clusters, using
a method described in detail by Coles, Davies \& Pearson (1996) and
which is similar (but not identical) to the well--known {\em genus}
statistic (reviewed by Melott 1990) and which has recently been applied
to cluster data by Rhoads, Gott \& Postman (1994). We next, in Section 5,
discuss an analysis based on percolation theory. The last of our three
approaches, presented in Section 6, 
is based on properties of a graph-theoretical construction
known as the minimal spanning tree, in conjunction with a set of
mathematical quantities intended to describe the shapes of pieces
of the trees obtained (Pearson \& Coles 1995). Each of the three
analyses we attempt is expected to perform better in some situations
than others, so in Section 7 we present an analysis of the 
statistical {\em power} of these tests at discriminating between
different models and between the models and the observed data.  
We also discuss the virtues of combining the various tests and show
the statistical significance of the results we obtain by combining
the different analyses into a composite test. We present our conclusions
in Section 8.

\section{The simulations}

\subsection{The Zel'dovich Approach}
The Zel'dovich approximation [ZA] (Zel'dovich 1970; Shandarin \& Zel'dovich
1989) is based on the assumption of laminar flow for the motion of a
self--gravitating non--relativistic collisionless fluid. Let $\bq$ be the
initial (Eulerian) position of a fluid element and $\br(\bq
,t)=a(t)\,\bx(\bq, t)$ the final position at the time $t$, which is related
to the comoving Lagrangian coordinate $\bx(\bq,t)$ through the cosmic
expansion factor $a(t)$. The ZA amounts to assume the expression
\be
\br(\bq,t)~=~a(t)\,\left[\bq+b(t)\,{\bf \nabla}_{\bq}\psi(\bq)\right]
\label{eq:za}
\ee
for the Eulerian--to--Lagrangian coordinate mapping. In equation (\ref{eq:za})
$b(t)$ is the growing mode for the evolution of linear density
perturbations and $\psi(\bq)$ is the gravitational potential, which is
related to the initial density fluctuation field, $\delta(\bq)$,
through the Poisson equation
\be
{\bf \nabla}^2\psi(\bq)~=~-{\delta(\bq)\over a(t)}\,.
\label{eq:poi}
\ee
As a result of the factorization of the $t$-- and $\bq$--dependence in the
displacement term of equation (\ref{eq:za}), the fluid particles move 
under this approximation along straight lines, with comoving peculiar velocity
\be
\bv(\bq,t)~=~\dot {\bx}(\bq,t)~=~\dot b(t)\,{\bf \nabla}_{\bq}\psi(\bq)\,.
\label{eq:vel}
\ee
Therefore, gravity determines the initial kick to the fluid particles
through eqs.(\ref{eq:poi}) and (\ref{eq:vel}), and afterwards they do not
feel any tidal interactions. Particles fall inside gravitational wells to
form structures, which however quickly evaporate. In this sense, the ZA
gives a good description of gravitational dynamics as far as particle
trajectories do not intersect with each other, while its validity breaks
down when shell--crossing occurs, and local gravity dominates. 

Coles, Melott \& Shandarin (1993) have shown that filtering out
the small--scale wavelength modes in the linear power--spectrum reduces the
amount of shell--crossing, thus improving the performance of the ZA. Melott,
Pellman \& Shandarin (1993) claimed that an optimal filtering procedure is
obtained by convolving the linear power--spectrum with the Gaussian filter
\be
W_G(kR_f)=e^{-(kR_f)^2/2}
\label{eq:gau}
\ee
(cf. Sahni \& Coles 1995).
The problem then arises of  choosing the filtering radius $R_f$ appropriately, 
in order to suppress shell--crossing as much as possible without
preventing genuine clustering to build up. 

Kofman et al. (1994) derived an analytical expression -- their equation  (7) 
-- for the average number of streams at each Eulerian point, $N_s$, as a function of
the r.m.s. fluctuation level of the initial Gaussian density field. 
We decided to choose $R_f$ for each model so that $N_s=1.1$. We found this
to be a reasonable compromise between smaller $N_s$ values, giving rapidly
increasing $R_f$ and high suppression of clustering, and larger $N_s$, at
which the ZA progressively breaks down. The resulting r.m.s. fluctuation
value corresponding to $N_s=1.1$ is $\sigma = 0.88$.

By adopting this implementation of the ZA, the main steps of our cluster
simulations are the following:

\begin{description}
\item[(a)] Convolve the linear power--spectrum with the Gaussian window
of equation (\ref{eq:gau}) and $R_f$ chosen as previously described.
\item[(b)] Generate a random--phase realization of the density field on
128$^3$ grid points for a cubic box of $L=320\hm$ aside.
\item[(c)] Move 128$^3$ particles having initial Lagrangian position on the
grid, according to the ZA. Each particle carries a mass of $4.4\times
10^{12} h^{-1}\,\Omega_{\circ}\, M_\odot$.
\item[(d)] Reassign the density and the velocity field on the grid through
a TSC interpolation scheme (e.g. Hockney \& Eastwood 1981) for the mass
and the moment carried by each particle.
\item[(e)] Select clusters as local density maxima on the grid according to
the following prescription. If $d_{cl}$ is the average cluster separation,
then we select $N_{cl}=(L/d_{cl})^3$ clusters as the $N_{cl}$ highest
density peaks. In the following, we assume $d_{cl}=40\hm$, which is
appropriate for the combined Abell/ACO cluster sample to which we will
compare our simulation results (see Section 3). Therefore, we will analyze a
distribution of 512 clusters in each simulation box,
with periodic boundary conditions.
\end{description}

\subsection{Dark Matter Models}
We ran simulations for six different models of the initial fluctuation
spectrum. For each model, we generate 5 random realizations, as a
compromise between estimating the cosmic variance reliably and keeping
the amount of data to be analysed within reasonable bounds.
All the models, except OCDM, are normalized 
to be consistent with the COBE measured
quadrupole of CMB temperature anisotropy (Bennett et al. 1994).
Since we are primarily interested in these simply as tests of the
method, we have not attempted to fine--tune the parameters of
each scenario in order to maximise its performance: the models
chosen are simply meant to represent the range of behaviours
of contenders for a viable model of structure formation.
The models we have considered are the following. 
\begin{description}
\item[(1)] The standard CDM model (SCDM), with $\Omega_\circ=1$, $h=0.5$ and
$\sigma_8= 1$ for the
r.m.s. fluctuation amplitude within a top--hat sphere of $8\hm$. This model
has already been excluded by independent analyses but we include here
for completeness and to see whether our pattern descriptors can
also successfully reject it.
\item[(2)] An open CDM model (OCDM), with $\Omega_0=0.2$ and $n=1$.
We have chosen to normalise this model to $b=1$, so that our results
in this paper can be compared with Plionis et al. (1995) and
Borgani et al. (1995); more detailed discussion of this model can
be found in (Coles \& Ellis 1994,1997; Ratra \& Peebles 1994,1995; Liddle
et al. 1996; Yamamoto \& Bunn 1996).
\item[(3)] A tilted CDM model (TCDM), with $n=0.7$ for the primordial
spectral index. Tilting the primordial spectral shape
from the scale--free one has been suggested in order to improve the CDM
description of the large--scale structure (e.g. Cen et al.
1992;  Tormen et al. 1993;
Liddle \& Lyth 1993; Adams et al. 1993; Moscardini et al. 1995).
\item[(4)] A low Hubble constant CDM model (LOWH), with $h=0.3$. 
Decreasing the Hubble constant has the effect of
increasing the horizon size at the equivalence epoch, thus pushing the
turnover of the spectrum to its scale--free form out to larger
scales (cf. Bartlett et al. 1994).
\item[(5)] A Cold + Hot DM model (CHDM), with $\Omega_{\rm hot}=0.3$ for the
fractional density contributed by the hot particles.
For a fixed large--scale normalization, adding a hot component has the 
effect of suppressing the power--spectrum amplitude at small wavelengths 
(e.g. Klypin et al. 1993).
Although the small--scale peculiar velocities are lowered to an adequate
level, the corresponding galaxy formation time is delayed so that such a
model is strongly constrained by the detection of high--redshift objects
(e.g. Ma \& Bertschinger 1994; Klypin et al. 1995; Borgani et al. 1997).
\item[(6)] A spatially flat, low--density CDM model ($\Lambda$CDM), with
$\Omega_{\circ} = 0.2$, $\Omega_{\Lambda} = 0.8$ for the cosmological
constant term (e.g. Efstathiou, Sutherland \& Maddox 1990;
Bahcall \& Cen 1992; Baugh \& Efstathiou 1993; Kofman, Gnedin \& Bahcall 1993;
Peacock \& Dodds 1994) and $\sigma_8=1.3$,
so as to be consistent with the two-year COBE results. 
\end{description}

The transfer functions for the above models have been taken from
Holtzman (1989), except that of LOWH, which is taken from Bond \&
Efstathiou (1984), with suitably chosen shape parameter $\Gamma = \Omega_oh
=0.3$. We note that the latter transfer function assumes the baryonic
component to be negligible, which is probably not accurate if
nucleosynthesis is correct, but this would only affect the shape of
the transfer function on small scales, below those we are interested in
here. All the model parameters are listed in Table \ref{t:dm}.

\begin{table}[tp]
\centering
\caption[]{The models. Column 2: the density parameter $\Omega_0$; Column 3:
the cosmological constant term $\Omega_{\Lambda}$; Column 4: 
the density parameter of
the hot component $\Omega_{\rm hot}$; Column 5: the primordial spectral
index $n$; Column 6: the Hubble parameter $h$; Column 7: the linear r.m.s.
fluctuation amplitude at $8\hm$ $\sigma_8$; Column 8: the filtering radius, 
$R_f$, in units of $\hm$, corresponding to $N_s=1.1$ for the level of 
orbit crossing.}
\label{t:dm}
\tabcolsep 5pt
\begin{tabular}{lccccccc} \\ \\
 Model & $\Omega_0$ & $\Omega_{\Lambda}$ & $\Omega_{\rm hot}$ & $n$ & $h$ &
$\sigma_8$ & $R_f$ \\ \\
 SCDM & 1.0 & 0.0 & 0.0 & 1.0 & 0.5 & 1.0 & 4.4 \\
 OCDM & 0.2 & 0.0 & 0.0 & 1.0 & 1.0 & 1.0 & 4.5 \\
 TCDM & 1.0 & 0.0 & 0.0 & 0.7 & 0.5 & 0.5 & 1.6 \\
 LOWH & 1.0 & 0.0 & 0.0 & 1.0 & 0.3 & 0.6 & 2.4 \\
 CHDM & 1.0 & 0.0 & 0.3 & 1.0 & 0.5 & 0.7 & 2.2 \\
 $\Lambda$CDM & 0.2 & 0.8 & 0.0 & 1.0 & 1.0 & 1.3 & 6.3 \\
\end{tabular}
\end{table}

It is worth making a specific point about the $\Lambda$CDM model we use here.
Strictly speaking, the amplitude of matter fluctuations required for
this model to be compatible with COBE is larger than can be treated with
great accuracy by our simulation method (Borgani et al. 1995). However,
we found in the course of this analysis that this feature of $\Lambda$CDM reveals
some interesting properties of the topological descriptors we use so,
rather than using an alternative model with a lower normalisation (c.f.
Borgani et al. 1995),
we will keep the model with a higher normalisation in this analysis.
In any case, as we shall show, changing the normalisation from $\sigma_8=1.3$
to, say, $\sigma_8=0.8$ does not significantly change the large--scale 
topological properties of selected clusters.

\section{The Cluster Sample}
We use the combined Abell/ACO $R\ge 0$ cluster sample, as defined in
Plionis \& Valdarnini (1995) and in Borgani et al. (1995). 
The declination limit between the northern (Abell) and southern 
(ACO; Abell, Corwin \& Olowin 1989) sample is dec $\ge -17^{\circ}$ while
both samples are limited in Galactic latitude by $|b|\ge 30^{\circ}$.

To take into account the effect of Galactic absorption, we assume
the usual cosecant law:
\begin{equation}\label{eq:obs1}
P(|b|) = \mbox{dex} \; \left[ \alpha \left(1 - \csc |b| \right) \right]
\end{equation}
with $\alpha \approx 0.3$ for the Abell sample (Bahcall \& Soneira 1983;
Postman et al. 1989) and $\alpha \approx 0.2$ for the ACO sample
(Batuski et al. 1989). The cluster--redshift selection function, $P(z)$, is
determined in the usual way (cf. Postman et al. 1989), by fitting
the cluster density, as a function of $z$. 
Cluster distances are estimated using the standard relation:
\begin{equation}
R~ =~ \frac{c}{H_{\circ} q_{\circ}^{2} (1 + z)} \left[q_{\circ}z + 
(1-q_{\circ}) (1-\sqrt{2 q_{\circ} z +1}) \right],
\label{eq:mat}
\end{equation}
with $H_{\circ} = 100 \; h \;$ km sec$^{-1}$ Mpc$^{-1}$ and
$q_{\circ} = \Omega_o/2$. Strictly speaking, equation (\ref{eq:mat}) holds only
for the case of a 
vanishing cosmological constant. Therefore, for a consistent comparison
with the simulation models, we should use different $R$--$z$ relations for
the Abell/ACO analysis. However, we verified that final results are
essentially independent of the choice of the $(\Lambda, \Omega_o)$
parameters used in the simulations. For this reason, in the following we
will present results for real data only based on assuming equation 
(\ref{eq:mat}) with $q_o=0.2$.

Note that due to the size of our simulations ($L=320$ $h^{-1}$ Mpc)
we will restrict our analysis within a sphere of radius 160 $h^{-1}$ Mpc. 
Within this volume our Abell/ACO cluster sample is complete
($P(z)\approx 1$) containing clusters all of which have 
measured redshifts. The Abell and ACO cluster number densities, 
corrected for galactic absorption according to equation (\ref{eq:obs1}) 
and within the present sample limits, are $\sim 1.7 \times 10^{-5}$ 
$h^{3}$ Mpc$^{-3}$ and $\sim 2.3 \times 10^{-5}$ $h^{3}$ Mpc$^{-3}$ 
respectively.
The higher space--density of ACO clusters is partly due to the
unique Shapley concentration (Shapley 1930; Scaramella et al. 1989), 
but a part is also due to systematic density differences between 
the Abell and ACO cluster samples which has been noted in  a number of studies
(cf. Plionis \& Valdarnini 1991 and references therein) and which could be 
attributed to the high sensitivity of the IIIa--J emulsion plates. 
In fact, excluding the small 
$b>30^{\circ}$ region of the ACO sample where the Shapley concentration lies
(corresponding to a solid angle of $\delta\Omega \approx 0.08 \pi$) 
lowers significantly the ACO cluster density ($\sim 1.9 \times 10^{-5}$ 
$h^{3}$  Mpc$^{-3}$).
Therefore the mean Abell/ACO cluster separation of our sample is 
$d_{cl} \approx 38 - 40 \hm$. 
 
In the following, we compare results based on the Abell/ACO sample with
those derived from our simulated cluster populations, 
selected to have a similar number-density,
sample volume and sky coverage. We have verified that variations in
$d_{\rm cl}$ of the order of the Abell--ACO difference, do not significantly
affect the resulting statistical properties.

It is important to stress that there is a possibility that these catalogues
may be contaminated by selection effects, perhaps due to the line-of-sight
projection effects (e.g. Sutherland 1988; but see Jing, Plionis \& Valdarnini 
1992). The effect
of such errors may be less pronounced on the `morphological' measures
of clustering we employ in this paper than on the quantities such as
the two--point correlation function used in previous work. Nevertheless,
the uncertain reliability of the catalogue, together with its relatively
small size, requires us to be circumspect when presenting our conclusions.

\section{Topology}
In this section we explore the behaviour of a topological characteristic
of the large-scale distribution of clusters, called the Euler--Poincar\'{e}
characteristic, for the different simulated data sets. For general background
material on topology, see Adler (1981) and Nash \& Sen (1983).

\subsection{Theory}
One of the commonly--used quantitative measures of clustering pattern
used in cosmology is the so-called {\em genus} statistic, described in
detail in Melott (1990) who gives  the genus $g$ of a solid object as
\begin{equation}
g\equiv (\mbox{no. holes}) - (\mbox{no. isolated regions}) + 1.
\label{eq:genus}
\end{equation}
This characteristic is generally applied to the observational data 
by first smoothing
them to form a continuous density field, $\delta$, 
and then locating the regions
where the smoothed field exceeds a given threshold density. Isodensity
surfaces thus define solid three-dimensional objects whose topology can
be defined in terms of the genus. One typically labels the threshold density
in the dimensionless form, $\nu$, defined as the number of standard deviations
of $\delta$ above the mean: $\delta=\nu\sigma$ (the mean value of
$\delta$ is zero by construction). 
One of the great advantages of the characteristic
$g$ is that, for a Gaussian density field in three dimensions, its mean
value per unit volume, $g_S$, as a function of $\nu$ can be obtained in a
simple closed form:
\begin{equation}
g_S = A(1-\nu^{2}) \exp (-\nu^{2}/2)\label{eq:g}
\end{equation}
(Doroshkevich 1970; Adler 1981; Bardeen et al. 1986;
Hamilton et al. 1986); the constant $A$ depends only
on the first and second moments of the power spectrum of $\delta$
and can be expressed in terms of the {\em coherence length} of the
random field, $\lambda_c$:
\begin{equation}
A=\frac{1}{4\pi^2 \lambda_c^3}
\end{equation}
where
\begin{equation}
\frac{1}{\lambda_c^{2}} \equiv \frac{<k^{2}>}{3},
\end{equation}
and
\begin{equation}
<k^{2}> \equiv \frac{\int_0^{\infty} P(k)k^4 dk}{\int_0^{\infty} P(k) k^{2}
dk}.
\end{equation} 
The dependence (\ref{eq:g}) 
means that all Gaussian fields produce the same
shape curve for $g_S(\nu)$ and that the amplitude can, in principle
at least, be used to determine properties of the power spectrum, $P(k)$,
relatively directly from the data. Note that the coherence length will
be determined both by the shape of the transfer function of the 
model in question and by the scale of smoothing adopted to produce
the continuous density field; see below.

The genus curve for
Gaussian fields is symmetric about the mean and positive for $|\nu|<1$,
indicating that threshold values around the mean give rise to
contour surfaces which are multiply connected. This is characteristic
`sponge' topology in which high density and low density regions
interlock. Non--Gaussian alternatives would be a `meatball' topology
in which isolated high density regions sit in a low-density background
and the mirror--image of this, a `swiss--cheese' topology.

The quantity $g_S$ is usually measured in practice by invoking the
Gauss--Bonnet theorem to relate it to the integrated curvature of
the contour surfaces; the algorithm CONTOUR3D is the standard tool
for performing this calculation on smoothed observational or
simulated data sets (Gott, Melott \& Dickinson 1986; Hamilton, Gott
\& Weinberg 1986; Melott, Weinberg \& Gott 1988;
Gott et al. 1989; Melott 1990; Moore et al. 1992; Vogeley et al. 1994). 
In a more recent paper, however,
Coles et al. (1996) have shown that a much more simple and
efficient topology-measuring algorithm can be developed from
ideas presented by Adler (1981). This algorithm basically
computes an approximation to the Euler--Poincar\'{e} Characteristic 
(EPC) $\chi$
of data defined on a grid or lattice in three dimensions. If the
genus is defined according to equation (\ref{eq:genus}) then $g=-\chi/2$
so that the curves of $\chi(\nu)$ and $g(\nu)$ are of the same
shape except for a sign. We shall concentrate on $\chi$ from now on.
The algorithm we use is 
explained in more detail in Coles et al. (1996), but
basically one constructs a three-dimensional framework of points, lines,
squares and cubes linking neighbouring points above the threshold
density contrast. If there are $P$ points, $L$ lines, 
$S$ squares and $C$ cubes
then
\begin{equation}
\chi\simeq P-[L+C]+S.
\end{equation}
Points are counted whether or not they belong to lines, squares
or cubes; lines are counted whether or not they belong to squares
or cubes; squares are counted whether or not they form part of
cubes. This calculation is a simple generalisation of the
2D equivalent which counts only points, lines and squares:
the 2D version has been explored in detail in (Coles 1988;
Coles \& Plionis 1991; Plionis, Valdarnini \& Coles 1992;
Davies \& Coles 1993; Coles et al. 1993). Alternative algorithms
are also discussed in Coles et al. (1996).

In order to define the excursion sets appropriately one needs to
smooth the initial point set with some kind of local averaging procedure.
Clearly the smoothing radius adopted must be greater than, or of the order
of the mean distance between points otherwise a continuous field is not
created. To implement our algorithm as described above we also need to
grid the data on a regular cubic lattice. The choices of grid resolution
and smoothing scale are user--defined quantities and must be chosen
in a pragmatic fashion. For example, the coherence of the density
field should not be too large compared with the sample volume, otherwise
edge effects dominate. A correction for edge effects is straightforward
for periodic boundaries, such as in our simulations but is less reliable
if there is a complicated boundary. One also wants the gridding to be
fine enough that each piece of the excursion set is sampled by
a sufficient number of cell points and that the ratio of the total
number of points in the sample volume to the number on the edges is large.
The smoothing scale adopted also depends on the number density of points
selected: for richer clusters we need a longer smoothing length, and this
may also affect the optimal choice of gridding. We discuss our final
choices below.

We finally remark that we prefer to plot the behaviour $\chi(\nu)$ as
a function
of $\nu$ as defined above in terms of the standard deviation of the density
fluctuations. Other authors (e.g. Melott 1990) prefer to plot a different
version of these curves which uses the volume fraction above the threshold
to calculate the effective value $\nu$ would have for the same volume fraction
of a Gaussian random field (using the error function). Any dependence
of the results on the one--point distribution of the fluctuations is
transformed away in this latter definition, so it has the
advantage of removing any effect of a monotonic
local bias (e.g. Kaiser 1984; Coles 1993)
on initially Gaussian fluctuations: the volume fraction remains 
the same in such a transformation, since excursion sets in the unbiased field
are mapped into the same sets in the biased field. The justification for
this is that one might be able to recover the topology of the initial
density field from that of a set of locally--biased mass tracers by 
exploiting this property. On the other hand, this
definition may conceal information about the form of the bias if it is 
non--local or non--monotonic. In the case we are interested in, the 
clusters are defined
as peaks of a non--linear and therefore non--Gaussian density field and
it is not clear what the effect of mapping back onto a Gaussian distribution
will have. We therefore feel that it is better not to attempt to remove
one--point information, as this may yield important clues about the biasing
of clusters of different density relative to the mass distribution. 
One expects,  for example,
that clusters with higher density would be more biased than those of lower
density and would therefore have a curve with a stronger apparent
meatball shift.

For reference, we should point out that related approaches to the
analysis of superclustering have been implemented recently. Rhoads,
Gott \& Postman (1994) have used the more standard `genus' algorithm
to study the topological properties of contour surfaces constructed
from the cluster distribution, including the different choice
of $\nu$ we described above. On the other hand, Kerscher et al. (1997)
have used a different mathematical approach, based on the so--called
Minkowski functionals, which incorporates as one of the descriptors  
a quantity analogous to the genus; for a further development of these
ideas, see Schmalzing \& Buchert (1997). These analyses are to some
extent similar in spirit to that which we present here, but there are
significant differences in both philosophy and implementation.

\subsection{Analysis}
In this section, we discuss only the comparison of our models with each
other and leave the detailed analysis of errors, confidence intervals and 
comparisons with the data until Section 6.

We performed a series of calculations of the EPC for the simulations with the
standard correction for periodic boundary conditions (Coles et al. 1996).
To get a feel for the effect of cluster selection on the strength
of the meatball distortion produced we have considered 
the distribution of all the density peaks identified on the grid, as
well as the distribution of the highest peaks, selected so as 
to produce clusters with a mean
spacing of $40 h^{-1}$ Mpc, comparable to the Abell/ACO catalogue.
\begin{center}
Figure 1
\end{center}
Results are shown in Figure 1 for the EPC characteristic $\chi(\nu)$.
We have chosen a Gaussian smoothing radius of $10 h^{-1}$ Mpc, and have binned
the smoothed field onto a $32^3$ grid. The curves shown are an
average over 5 realizations of the model concerned, with error bars
representing the standard deviation over this ensemble.
Notice the slight asymmetry compared
to the expected behaviour for a Gaussian field and
the apparently anomalous shape of the $\Lambda$CDM, which appears to be
very different to the other models. One should be suspicious that this
difference might be due to the fact that this model is more highly
evolved (i.e. it has a higher value of $\sigma_8$) than the others
and the strange topological behaviour is simply due to the fact
that our Zel'dovich simulation 
method is behaving badly for this model. In fact,
this is not the case. We applied the same test to a less 
evolved $\Lambda$CDM model 
($\sigma_8=0.8$; cf.  Borgani et al. 1995), which was
demonstrated 
to be evolved quite accurately by our
Zel'dovich technique, but found a graph of the EPC which matched
the more evolved model closely in shape. It seems therefore that
the more pronounced behaviour of the $\Lambda$CDM is actually connected with
intrinsic properties of the model,
rather than  with any limitation
of our simulation method.
\begin{center}
Figure 2
\end{center}
Figure 2 shows analogous results to Figure 1, but
for clusters selected to resemble Abell/ACO clusters; we have
chosen a Gaussian smoothing radius of $30 h^{-1}$ Mpc, and have binned
the smoothed field onto a $32^3$ grid. Notice the slight asymmetry compared
to the expected behaviour for a Gaussian field. The curves shown are again an
average over 5 realizations. Note the reduction in amplitude, due to the
increase in smoothing length resulting in an increased coherence length,
and a more drastic meatball effect as a consequence of the bias: the
point of minimum $\chi$ is moved further to the left than in Figure 1.
Notice also the substantially greater noise, even after averaging over
5 simulations. 

It is important also to note the similarity of the curves for different
models in Figure 2 compared to the clear systematic variations of the
curves with model in Figure 1. This shows that the dominant effect
on the behaviour of the EPC in the cluster--selected samples is that
of thresholding rather than in difference in the amount of evolution
or in the shape of the initial power spectrum. In particular,
notice that the shape of the EPC curve for $\Lambda$CDM, although anomalous
in Figure 1 is consistent with the other models when only selected clusters 
are used. There are nevertheless
residual differences due to these other factors and, as we shall show
later, they do allow discrimination between the models with some
degree of statistical confidence.

\section{Percolation}

\subsection{Theory}
The use of percolation methods (e.g. Stauffer \& Aharony 1992; 
Isichenko 1992), which have been borrowed by
cosmologists from condensed matter physics, to study  aspects of
galaxy clustering dates back to Zel'dovich (1982) and Shandarin (1983), 
and for
quantifying observed properties of galaxy clustering to
Einasto et al. (1984); see also Zel'dovich, Einasto \& Shandarin (1982). 
Initially, the method used was based on the
idea of ``decorating'' each point (galaxy or cluster) in a point set with a
sphere of some radius and determining the point at which these spheres
overlap to `percolate' the entire set. Suppose we have $N$ objects in
a cubic sample of side $L$. The mean object--object separation is 
defined to be $l=LN^{-1/3}$. One (notionally)
draws a sphere of diameter $d=bl$, where $b$ is dimensionless, around each 
point and determines $L_p(b)$, the maximum distance that can be 
traversed while still remaining within such spheres. The spheres around
neighbouring points may, of course, overlap with each other. When $L_p(b)=L$
then the set is said to percolate and the critical value of $b=b_*$ is
called the percolation parameter. For a uniform set of points on a 
Cartesian lattice, $b_*=1$, while if the points are distributed along
lines or sheets, $b_*<1$. This simple indicator of clustering, using
only $b_*$ to quantify the connectivity of structures has not been
altogether successful (Bhavsar \& Barrow 1983; Dekel \& West 1985;
Dominik \& Shandarin 1992), partly
due to sensitivity to sample selection parameters but mainly due
to the inadequacy of encoding in one numerical
quantity (i.e. $b_*$) all the properties of the transition
from the unpercolated distribution to a percolated one.

To remedy this shortcoming, and to keep the analysis as comparable as
possible to that of the preceding Section on the EPC, we adopt a
more sophisticated approach, as described by Klypin \& Shandarin (1993);
see also Mo \& B\"{o}rner (1990), de Lapparent et al. (1991),
Yess \& Shandarin (1996), Sathyaprakash, Sahni \& Shandarin (1996)
and Sahni, Sathyaprakash and Shandarin (1997). In this
approach, we consider the application of percolation techniques to
a cubic lattice on which, according to some density threshold criterion,
cells are labelled as either `filled' or `empty'. Once a cell is
so labelled, clusters of cells are identified. A cluster can be defined
as a connected neighbourhood of cells, where cells are connected if they
either share a common side (not an edge or corner) or have a neighbour
which is connected according to the previous criterion 
(i.e. `friends--of--friends'). From this definition of clusters
(actually, in this context, it would be more accurate to call them
`superclusters'), we define the size of a cluster to be the number of
cells in the cluster. An infinite cluster is so called if it connects
antipodal sides of the cubic lattice and the critical threshold
level $\rho_c$ is the threshold level at which the first infinite cluster
is formed. At $\rho_c$ the system can be thought of as undergoing a 
kind of phase transition, from an unconnected to a connected state.
If the cubic lattice has side $L$ then the filling factor of the lattice
is defined to be simply the fraction of cells labelled as ``filled'', this
yields the probability $p$ of a randomly--selected cell being filled. We
then define the {\em multiplicity function}, $n(\eta)$, to be the average
number density of clusters of size $\eta$. From this a cell can be thought of
as being in one of three states: empty (with probability $1-p$);
member of a finite (non--spanning) cluster with probability
\begin{equation}
p_+=\sum_{\eta} \eta n(\eta);
\end{equation}
or part of an infinite cluster with probability 
\begin{equation}
p_\infty =  \eta_{\rm max}/L^{3}\, .
\end{equation}
We then employ two sample statistics derived from these considerations. First is
the fraction of cells belonging to infinite clusters
in a given sample, which is an estimator of $p_\infty$
and which we shall call $\mu_\infty$. We define the second parameter
$\mu^{2}$ to be the (weighted) mean square size of all clusters excluding
the largest one:
\begin{equation}
\mu^{2} = \frac{\sum_\eta \eta^{2} n(\eta)}{L^{2/3}\sum_\eta n(\eta)}~;
\end{equation}
the factor of $L^{2/3}$ is simply to scale results for different $L$
onto each other more simply (Klypin \& Shandarin 1993).

\subsection{Analysis}

For the percolation analysis of the cluster simulations we have followed
a similar approach to that of the EPC. That is to say, we have analysed
samples containing both ``all'' clusters and those selected according to
the Abell/ACO number--density criteria. We have analysed the simulations
according to a variety of different grid resolutions but, for conciseness
and to show only the most pertinent trends, we describe only those results
for the same smoothing and gridding onto
a cubic mesh as in Section 4; as before we define the density
threshold $\rho_c$ in terms of the number of standard deviations from the
mean density. 
\begin{center}
Figures 3 \& 4.
\end{center}
Figure 3 shows the behaviour of $\mu_\infty$ for all the clusters in each
model, while Figure 4 shows the same plot for the Abell/ACO selected samples.
Notice that the fraction of points in the infinite cluster is generally
unity for low thresholds and zero for high thresholds independent of
the model, as expected. The transition from one limit to the other, however,
takes place at different threshold levels for the different models: it
begins at around $\nu\simeq -2$ for the `all data' samples and at around
$\nu=-1$ for the selected clusters; the transition is more
rapid in the latter case. This must again be mainly
influenced by the shape of the one--point distribution of the density fields in the two
cases rather than differences in the amount of evolution or in the shape
of the primordial fluctuation spectrum. As was the case for the EPC
curves, it is not obvious to the uneducated eye whether the curves
which are averages over 5 simulations, reveal significant departures
from one model to another: we shall discuss this in more detail later.
\begin{center}
Figure 5
\end{center}
Now we turn to the behaviour of $\mu^{2}$ displayed in Figure 5.
For brevity we show only
for the Abell/ACO type clusters and using a $32^3$ grid as before. 
Notice that the largest mean size
of cluster is formed at thresholds between $\nu=-1$ and $\nu=0$ which is
where the topology also indicates the highest degree of multiple connectivity.
There is an apparent `glitch' in the case of the LOWH model displayed in the
right--hand panel, but this is simply due to the fact that there are two
effectively infinite clusters in this case: the pattern percolates in two
directions at low thresholds and only one of the infinite clusters
is removed in the averaging procedure. Other than this one feature
(which occurs with a small but non--negligible probability) the
distributions are visually similar for all models.

At this stage we simply remark that differences between the models
appear to be less pronounced in terms of these two statistics than they
do when we look at the topological EPC descriptor. This conclusion is
not altered if one uses different gridding parameters; one simply
probes the connectivity on a different scale. A more detailed
analysis of the discrimination achieved between these models using these
descriptors is deferred until Section 7.

\section{Minimal Spanning Trees and Structure Functions}

In this section we describe the results of applying a test proposed
by Pearson \& Coles (1995) to these simulations. This test was originally
suggested as a means to quantify the shape (i.e. filament or sheet--like
geometry) of galaxy clustering, rather than its topology (i.e. connectivity)
which was the case in the previous two descriptors. Rather than
smoothing the data, one constructs the Minimal Spanning Tree (MST)
of the point set and then quantifies the shape of pieces of the tree
using a set of three shape parameters suggested by Babul \& Starkman (1992);
see also Luo \& Vishniac (1995) and Dav\'{e} et al. (1997).
We should say at the outset that the {\em shape} of superclustering
is expected to be much more poorly defined than that of galaxy clustering,
both because the cluster distribution is dominated by small number statistics
(which one ameliorates in the topology analysis by smoothing) and
because the formation of structures of a well--defined dimensionality
is not expected on the very large scales (which are
evolving in a quasi--linear fashion) probed by clusters. One might
expect the performance of any shape statistic therefore to be
rather poorer than a topological descriptor. This will, in fact, be
what we find.

\subsection{Theory}
Full details of our method are described in Pearson et al. (1995);
we simply define our notation here. The MST (Ore 1962; Gower \& Ross 1969;
Zahn 1971) is derived
from graph theory and is a construct that (uniquely) connects a
set of $N$ points (`nodes') with $N-1$ straight lines (`edges')
in such a way that the sum of the edge lengths is a minimum and there
are no closed circuits in the graph thus formed. For applications
of this construction to galaxy clustering problems, see
(Barrow, Bhavsar \& Sonoda 1984,1985; Bhavsar \& Ling 1988a,b; Plionis, 
Valdarnini \& Jing 1992; Krzewina \& Saslaw 1996). For an interesting
discussion of the relationship between the MST and the percolation
approach used in the previous section, see Bhavsar \& Splinter (1996).

Once the tree is constructed
it can be separated by removing all the edges that exceed a specified
cutoff length. As is usual, we define the cutoff length as a
multiple of the mean edge length of the MST so as to delete chance
linkages. After separation, the MST will fall into a number of
disjoint trees whose properties can be further explored individually.
Pearson \& Coles (1995) showed how separation can be used to
enhance structures relative to surrounding noise.

The MST is a construction rather than a statistic so in order to
use it to describe galaxy clustering we have to quantify the shape
of the tree(s). We have chosen to use the structure functions
$S_1$, $S_2$ and $S_3$ defined originally by Babul \& Starkman (1992).
These are calculated by first defining the moment of inertia tensor
around the centre of mass of each piece of the separated tree. The
eigenvalues of this tensor are used to define quantities $S_i$ such
that $0\leq S_i\leq 1$ and  $(S_1, S_2, S_3)=(0,0,1)$ for a spherical
distribution, $(S_1, S_2, S_3)=(0,1,0)$ for a flat sheet, and
$(S_1, S_2, S_3)=(1,0,0)$ for a straight filament. The functions
are further designed to fall away rapidly from unity as a structure
deviates from the shape specified by a particular value of $S_i$.
One can then look, for example, at the distribution of $S_i$ values
over the pieces of the tree; see Pearson \& Coles (1995) for further
details.

\subsection{Analysis}
The analysis we carry out here follows that in Pearson \& Coles (1995)
which was applied to the simulations of Borgani et al. (1994) and which
showed that this method could discern differences between the models
presented in that paper. For each data set we first construct the MST,
and then separate the tree as outlined above. The values of $S_i$
are then calculated for each separate piece of the tree. In Pearson
\& Coles (1995), it was found that the structure functions $S_1$ and
$S_2$ were close to zero for the Borgani et al. (1994) simulations
indicating the lack of any obvious filamentary or sheet--like pattern.
We again found this for the newer simulations.
The $S_3$ statistic, however, was shown to have an interesting behaviour,
and we therefore look at it further here.
\begin{center}
Figure 6
\end{center}
Figure 6 shows the distribution of $S_3$ values for all the
simulated clusters and for a separation length equal to $F\bar{x}$,
where $\bar{x}$ is the mean edge length and the results are integrated
over all values of $F$ for simplicity of presentation; the trends with
$F$ are quite consistent in the different models. The TCDM and
LOWH models have a more peaked distribution of $S_3$ values than the
other models, while the $\Lambda$CDM model is much broader but has a lower
average. Roughly speaking, this means that the distribution 
in the $\Lambda$CDM case is less spherical than in the others which is consistent
with the greater amount of dynamical evolution on large scales 
in this model than in the other cases, and which is also
evident in the analysis we performed in Section 4.

We have also looked at the differences in the number of trees formed
at each model at different edge cutoff lengths (not shown). Again, the
$\Lambda$CDM model stands out because it has the greatest distribution of
values over which trees form: all curves peak at around $F=1.25$ and
fall off rapidly as $F$ is increased.

Our most disappointing result, however, is that when we select
clusters with a mean spacing of $40 h^{-1}$ Mpc, we find that the
MST produces insufficient trees to proceed with the analysis on simulation
volumes of this size. When the tree is constructed and separation performed
with any reasonable value of $F$, one simply gets trees containing
only one node. We are therefore unable to use this descriptor for
more detailed tests of selected clusters. As we feared, small number effects
prevents us using this method for samples as sparsely sampled
and within such a small volume as our simulations.

The conclusion of this section is, therefore, that while the MST
method proposed by Pearson \& Coles (1995) can indeed discriminate
between different underlying distributions, the shot--noise
associated with clusters, and the lack of a pattern with a specific
dimensionality, seems to pose insuperable problems for testing the cluster
distribution using data sets of size comparable to those we
have used in this study.

\section{Statistical Tests}
The results we have discussed so far have been displayed for visual
interpretation only, and without a detailed study of the errors and
resulting confidence limits. We have shown results only for
the theoretical simulations,  and not for the real cluster catalogue.
Now there are two main tasks one might be interested in setting for
clustering descriptors of the kind we have discussed so far in this
paper. One is to indicate differences between the pattern displayed
by the different models so one can understand the  impact of
initial conditions and evolution on the clustering pattern. The
second task is to test specific models against real data to find out
whether they are compatible.

What we have done, therefore, is to recalculate the EPC and percolation
statistics for the Abell/ACO sample described in Section 3 (since the
MST method performs so poorly for the idealised samples, we do not
discuss it further in this section: it is even worse when applied to
the real catalogue, which is smaller). 
To obtain maximum discriminatory
power we do not restrict ourselves to one set of gridding or smoothing
parameters: we use a bank of results for $16^3$, $32^3$ and $64^3$ grids
and for four different choices of smoothing length: $10h^{-1}$, $20h^{-1}$,
$30h^{-1}$ and $40h^{-1}$ Mpc and for an ensemble of 5 simulations to calculate
significance levels for differences in behaviour of these descriptors
(i) between the models and (ii) between the models and the data.
We constructed distributions for each of the quantities involved
in the analyses we have described, and then used a Monte-Carlo test
using the  Kolmogorov-Smirnov statistic to find the fraction of times that
the distributions were found to be different. This then yields
a robust estimate of the statistical significance of differences
between the models under each of these descriptors.
 We looked at this question
for each descriptor separately and then, at the end,
when all descriptors are used in concert.
\begin{center}
Figure 7
\end{center}
The need for such a detailed statistical study is demonstrated by the form
of Figure 7 which shows the results obtained for the EPC smoothed and gridded
as in Section 4 for the Abell/ACO sample and for samples extracted from the
simulations according to the same selection criteria (i.e. radial
distribution and sky coverage). It is by no means
obvious whether there are any systematic departures, although some of the
model curves appear to be discrepant. Merely plotting error bars on this
curve would not help much as differences in the shape are more important
than differences in the amplitude.

We now display the results of this procedure in a series of tables
which all have the same format: along the downward diagonal we see
the significance level of departures of the simulated Abell/ACO
clusters (i.e. clusters selected according to the same
criteria as the Abell/ACO sample) 
against the real data; above the diagonal shows the significance
level of departures of models from each other based on the 
properties of all clusters in the simulation; below the diagonal
we have discrimination between models based on the properties
of clusters selected with a mean spacing of $40h^{-1}$ Mpc
(but still in a cubic volume). To give an example, in Table 2,
we see that the $\Lambda$CDM model disagrees with the Abell/ACO data
at a 42\% confidence level; while it is different from OCDM
at the 91\% level if all data are included and different to
CHDM at the 66\% level if only selected clusters are used.

\begin{table}[tp]
\centering
\caption[]{Power of EPC as a discriminator. The diagonal (in boldface) shows
the comparison with the real Abell/ACO data, above right of the diagonal
shows discrimination between models using 
the distribution of all the density peaks,
below left shows discrimination between models using selected
clusters only.}
\label{t:epc}
\tabcolsep 5pt
\begin{tabular}{ccccccc}
 & CHDM & OCDM & SCDM & $\Lambda$CDM & LOWH & TCDM\\
CHDM & {\bf 0.92} & 0.58 & 0.17 & 1.00 & 0.58 & 0.75\\
OCDM & 0.33 & {\bf 0.92} & 0.25 & 0.91 & 0.75 & 0.91\\
SCDM & 0.50 & 0.83 & {\bf 0.92} & 1.00 & 0.83 & 0.91\\
$\Lambda$CDM & 0.66 & 0.58 & 0.91 & {\bf 0.42} & 1.00 & 1.00\\
LOWH & 0.50 & 0.91 & 0.91 & 0.91 & {\bf 1.00} & 0.23\\
TCDM & 0.50 & 0.66 & 0.09 & 0.75 & 0.00 & {\bf 1.00}\\
\end{tabular}
\end{table}
Table 2 shows the results for the EPC only. Looking first at the
diagonal reveals that all but the LOWH and TCDM models are
consistent with the Abell/ACO data within 95\% confidence. The
best fit is $\Lambda$CDM. The ability of the method to discriminate
between models is variable and can be very poor if only the
selected clusters are used: for example, LOWH and TCDM appear
identical in this case.
\begin{table}[tp]
\centering
\caption[]{Power of $\mu_{\infty}$ as a discriminator. The diagonal 
(in boldface) shows
the comparison with the real Abell/ACO data, above right of the diagonal
shows discrimination between models using 
the distribution of all the density peaks,
below left shows discrimination between models using selected
clusters only.}
\label{t:mui}
\tabcolsep 5pt
\begin{tabular}{ccccccc}
 & CHDM & OCDM & SCDM & $\Lambda$CDM & LOWH & TCDM\\
CHDM & {\bf 0.92} & 0.33 & 0.25 & 0.66 & 0.33 & 0.91\\
OCDM & 0.33 & {\bf 0.92} & 0.91 & 0.75 & 0.66 & 0.83\\
SCDM & 0.09 & 0.50 & {\bf 1.00} & 0.66 & 0.50 & 0.83\\
$\Lambda$CDM & 0.66 & 0.25 & 0.75 & {\bf 0.50} & 1.00 & 1.00\\
LOWH & 0.17 & 0.50 & 0.17 & 0.83 & {\bf 1.00} & 0.33\\
TCDM & 0.25 & 0.41 & 0.17 & 1.00 & 0.00 & {\bf 0.92}\\
\end{tabular}
\end{table}
Table 3 shows the results for $\mu_\infty$ only. This
appears to rule out both SCDM and LOWH at the 95\% level, while $\Lambda$CDM
is again the best fit. Although the discrimination is again variable,
and this statistic does not simply track the behaviour of the EPC test,
the average power of discrimination is somewhat lower for this statistic
than for the EPC.
\begin{table}[tp]
\centering
\caption[]{Power of $\mu^{2}$ as a discriminator. The diagonal 
(in boldface) shows
the comparison with the real Abell/ACO data, above right of the diagonal
shows discrimination between models using 
the distribution of all the peaks,
below left shows discrimination between models using selected
clusters only.}
\label{t:mu}
\tabcolsep 5pt
\begin{tabular}{ccccccc}
 & CHDM & OCDM & SCDM & $\Lambda$CDM & LOWH & TCDM\\
CHDM & {\bf 1.00} & 0.66 & 0.66 & 0.91 & 0.91 & 1.00\\
OCDM & 0.58 & {\bf 1.00} & 0.50 & 1.00 & 1.00 & 1.00\\
SCDM & 0.58 & 0.75 & {\bf 1.00} & 0.75 & 0.91 & 0.91\\
$\Lambda$CDM & 0.66 & 0.83 & 0.50 & {\bf 1.00} & 1.00 & 1.00\\
LOWH & 0.73 & 0.83 & 0.83 & 0.66 & {\bf 1.00} & 0.91\\
TCDM & 0.83 & 0.66 & 0.75 & 0.66 & 0.50 & {\bf 1.00}\\
\end{tabular}
\end{table}
Table 4 shows the behaviour of $\mu^2$. In terms of this
statistic, all models appear to
be discrepant with the data. This may however be due to the fact
that for these small samples only a small number of `clusters'
are involved in the calculation of equation (15) and the results
are therefore oversensitive to fluctuations from sample
to sample. Our power tests are an attempt to quantify the
robustness of the statistic to fluctuations of this type,
but if the variation over the ensemble is too large they
will not be reliable. It is also possible that boundary
effects dominate the behaviour of this quantity for the
real data because these interact in a different way
in the percolation analysis than in the EPC analysis. 
In any case, there certainly seems to  be stronger
discrimination between models with $\mu^{2}$ than with 
$\mu_{\infty}$
alone.

Since the  tests are not overwhelmingly powerful on an individual
basis, we look at the results of combining a battery of these
three into one `supertest' which would make use of any complementarity
that exists in these descriptors.
\begin{table}[tp]
\centering
\caption[]{Power of all tests combined into a single discriminator. 
The diagonal (in boldface) shows
the comparison with the real Abell/ACO data, above right of the diagonal
shows discrimination between models using 
the distribution of all the density peaks,
below left shows discrimination between models using selected
clusters only.}
\label{t:all}
\tabcolsep 5pt
\begin{tabular}{ccccccc}
 & CHDM & OCDM & SCDM & $\Lambda$CDM & LOWH & TCDM\\
CHDM & {\bf 0.94} & 0.53 & 0.36 & 0.86 & 0.61 & 0.89\\
OCDM & 0.39 & {\bf 0.94} & 0.28 & 0.88 & 0.80 & 0.91\\
SCDM & 0.39 & 0.69 & {\bf 0.97} & 0.81 & 0.75 & 0.89\\
$\Lambda$CDM & 0.67 & 0.56 & 0.72 & {\bf 0.64} & 1.00 & 1.00\\
LOWH & 0.47 & 0.75 & 0.36 & 0.81 & {\bf 1.00} & 0.50\\
TCDM & 0.53 & 0.58 & 0.33 & 0.83 & 0.17 & {\bf 1.00}\\
\end{tabular}
\end{table}
Table 5 shows the effectiveness of combining all the tests into one.
Notice that LOWH, SCDM and TCDM are all excluded with at least 95\%
confidence, but that the best discrimination that can be achieved between
the models, though better than in the previous tables, is generally
less than 95\%.

\section{Discussion and Conclusions}
In the Introduction to this paper, we stressed that this analysis was
to be treated as exploratory because there were reasonable grounds
to doubt the quality of present clustering data and that looking
for geometrical signatures of the pattern of superclustering was
in any case difficult because of the extreme rareness of rich clusters
and the consequent sparse sampling and shot-noise this implies.

Nevertheless, as a guide to the results one might expect from larger
and better controlled cluster samples the results we have obtained
are extremely encouraging, at least for some of the tests we
have used. Although this optimism is largely based on results from
simulations which may be reasonably argued to be much `cleaner' than
real data are likely to be, our results show at least that there
are perceptible differences between these models on large--scales
and that these do in principle allow one to discriminate between them
using shape- and topology-based descriptors.

For our topological analysis, based on the EPC, clear differences
emerge between the models. One has to be a little careful here, however,
because the form of the statistic we use actually contains information 
about the one--point distribution function of the objects, because
of the choice of threshold parameter $\nu$. Remember also that the amplitude
of the EPC curve is related to the coherence length of the density field
and that this is simply derived from the power spectrum. Comparing
the trends we see in the EPC analysis with the trends of the
one--point distribution found in an analysis of the same models
by Borgani et al. (1995) together with the coherence lengths of the
initial power spectra, shows that the behaviour of the EPC for
different simulations can, roughly speaking, be `explained' in terms
of these other descriptions. Although differences therefore show up
between the models, they are largely the same as the differences
one finds in non--topological descriptors. One would be justified
therefore in saying that this descriptor does not add very much:
it just provides a different way of seeing differences in one and
two--point information. Nevertheless, folding such information
in with the topology (which is in any case very easy to measure)
does seem to provide a simple methodology for discriminating between
models which does not require the computation of power-spectra and
distribution functions and may in any case incorporate at least
some extra information than these quantities do.

On the other hand, the topology of the Abell/ACO
data does not display the same kind of EPC graph that one would expect
by looking at the results of Plionis et al. (1995) and Borgani et al. 
(1995) and assuming it follows the same trends as our models. 
This may be telling us that the Abell/ACO is essentially
different to all of the models we have looked at in this paper,
which in turn may mean that either all the models are incorrect
or that there is something suspicious about the catalogues or the
way we have interpreted them. In particular, the effects of
redshift selection, galactic extinction and the differences
in number density between the Abell and ACO catalogues introduce
some uncertainty into our conclusions.

The one model that does have a topological description in reasonable
accord with the Abell/ACO data is the $\Lambda$CDM model, a result which agrees
with the results of Kerscher et al. (1997) (although the model they
used had a rather smaller value of $\Omega_\Lambda=0.65$ than the
model we have used here). This model also survives the tests
described in Borgani et al. (1995), but there was uncertainty
attached to that analysis because of the possibility of
that model being too strongly clustered to be adequately
described by the Zel'dovich approximations. We have shown
that this extra evolution does not influence the behaviour
of the EPC to any significant extent and the claim that this
model can reproduce the behaviour of Abell/ACO in terms
of topology and low-order moments therefore stands up to
scrutiny. This, of course, still admits the possibility that
this is telling us more about problems with the catalogue than
about the real distribution of overdensities.

The performance of our percolation test depends strongly
on the kind of statistic one extracts from the percolated
set. If one looks only at the statistic $\mu_\infty$ then
the power of discrimination is mediocre, but this rises
strongly if one uses $\mu^2$ instead or together with $\mu_\infty$.

The one disappointment of this analysis is the performance of
the MST/shape functions we introduced in Pearson \& Coles
(1995). Although they do perform well for relatively well--sampled
distributions, we were unable to get useful results for any
of the simulated samples of clusters. The application of this statistic,
at least in the form we have used it here, is not recommended
for extremely sparsely-sampled distributions like those of Abell
clusters.

Our final conclusion, however, is that topological and geometrical
descriptors (of which we have studied only three) are at least in principle
capable of diagnosing differences between very sparsely-sampled distributions
in a fashion which is quite independent of the one- and two-point
statistics which are more familiar in the cosmological community.
With the arrival of larger and better controlled samples of galaxy
redshifts and the cluster catalogues which will accompany them, clustering
data will not only be more amenable to this type of analysis, they will
also {\em require} such an approach if one is to extract as much information
as possible.

\section*{Acknowledgments}
RCP receives a PPARC research studentship. PC is a PPARC Advanced
Research Fellow.  LM  
thanks the Italian MURST for partial financial support. 
MP has been supported by an
EEC {\em Human Capital and Mobility} fellowship for some of
the period while this project was in progress.
This work has also been partly supported by funds originating
from the {\em EC} Human Capital and Mobility Network (Contract Number
CHRX--CT93--0129).

\newpage
\section*{Figure Captions}

\vspace{0.25cm}
\noindent
{\bf Figure 1.} Results for the EPC ($\chi$) as a function of density threshold,
$\nu$, expressed in standard deviations from the mean, for all models
described in the text using all peaks found in the simulations to define the
structure. Panel (a) shows CHDM (solid lines), $\Lambda$CDM (dotted) and LOWH
(dashed); Panel (b) shows OCDM (solid), SCDM (dotted) and TCDM (dashed).
Error bars are shown in (a) for CHDM and in (b) for OCDM only to avoid
crowding the plot. The vertical scaling is arbitrary, but is identical
for all the models.

\vspace{0.25cm}
\noindent
{\bf Figure 2.} Results for the EPC ($\chi$) as a function of density threshold,
expressed in standard deviations from the mean, $\nu$ for all models
described in the text using  clusters selected in the simulations
so as to have a fiducial mean spacing of $40 h^{-1}$ Mpc. 
Panel (a) shows CHDM (solid lines), $\Lambda$CDM (dotted) and LOWH
(dashed); Panel (b) shows OCDM (solid), SCDM (dotted) and TCDM (dashed).
Error bars are shown in (a) for CHDM and in (b) for OCDM only to avoid
crowding the plot. The vertical scaling is arbitrary, but is identical
for all the models.

\vspace{0.25cm}
\noindent
{\bf Figure 3.} Results for the percolation statistic $\mu_\infty$
for all models described in the text using all the peaks found in the
simulation to define the structure. 
Panel (a) shows CHDM (solid lines), $\Lambda$CDM (dotted) and LOWH
(dashed); Panel (b) shows OCDM (solid), SCDM (dotted) and TCDM (dashed).
Error bars are shown in (a) for CHDM and in (b) for OCDM only to avoid
crowding the plot. 

\vspace{0.25cm}
\noindent
{\bf Figure 4.} Results for the percolation statistic $\mu_\infty$
for all models described in the text using selected clusters only,
as in Figure 2.Panel (a) shows CHDM (solid lines), $\Lambda$CDM (dotted) and LOWH
(dashed); Panel (b) shows OCDM (solid), SCDM (dotted) and TCDM (dashed).
Error bars are shown in (a) for CHDM and in (b) for OCDM only to avoid
crowding the plot.

\vspace{0.25cm}
\noindent
{\bf Figure 5.} Results for the percolation statistic $\mu^2$
for all models described in the text using selected clusters only,
as in Figure 2. Panel (a) shows CHDM (solid lines), $\Lambda$CDM (dotted) and 
LOWH (dashed); Panel (b) shows OCDM (solid), SCDM (dotted) and TCDM (dashed).
Error bars are shown in (a) for CHDM and in (b) for OCDM only to avoid
crowding the plot.

\vspace{0.25cm}
\noindent
{\bf Figure 6.} Integrated distribution of the shape-space statistic
$S_3$ for all clusters selected and for all models. The distributions
are obtained by co-adding distributions for various values of $F$,
as described in the text. Panel (a) shows CHDM (solid lines), 
$\Lambda$CDM (dotted) and LOWH
(dashed); Panel (b) shows OCDM (solid), SCDM (dotted) and TCDM (dashed).
Error bars are now shown, as the results come from co-adding all the
simulation results.

\vspace{0.25cm}
\noindent
{\bf Figure 7.} Results for the EPC as a function of density threshold,
expressed in standard deviations from the mean, for all models described
in the text. Samples were extracted according to the same selection
criteria as the Abell/ACO sample which is also shown for comparison.
The figures show: (a) CHDM; (b) $\Lambda$CDM; (c) TCDM; (d) LOWH; (e)
OCDM; (f) SCDM; appropriate error bars are drawn on these curves.
The heavy solid line in each plot shows the corresponding results
for the Abell/ACO catalogue.
The noisiness of these curves demonstrates the need for careful
statistical assessment of the discriminatory power.

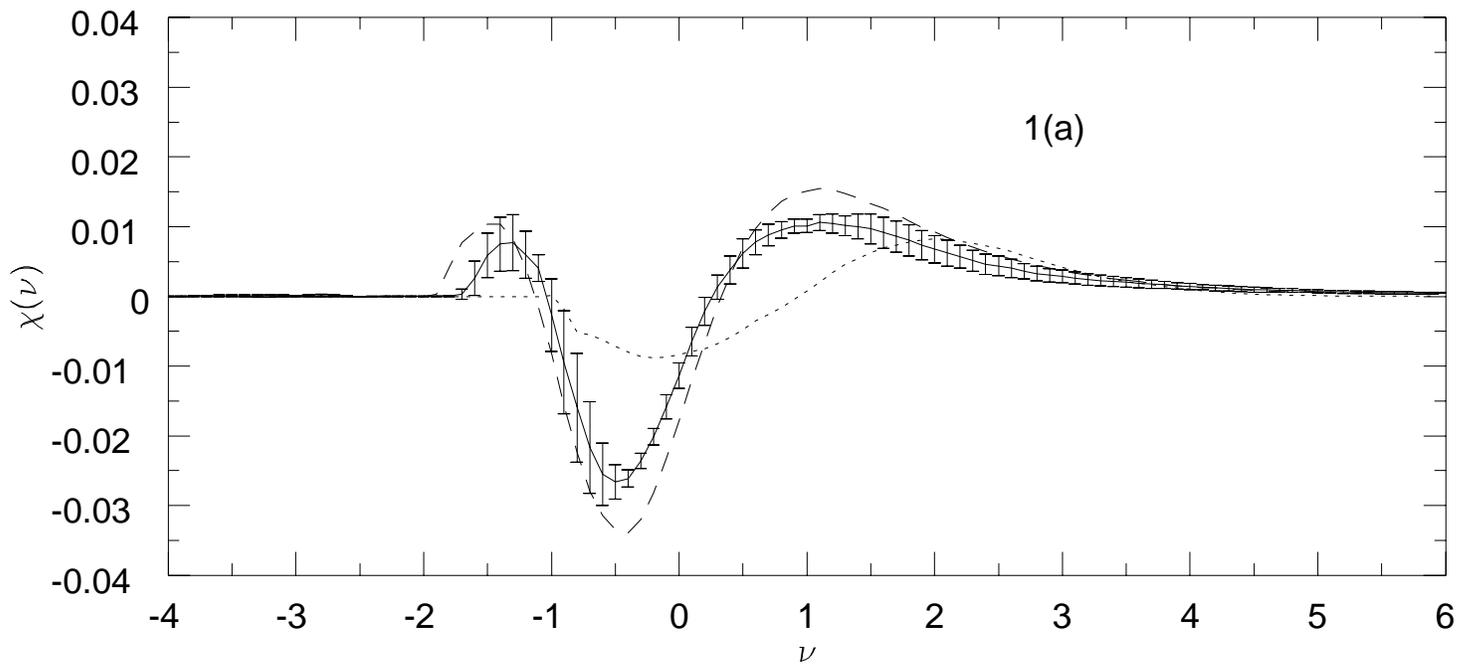
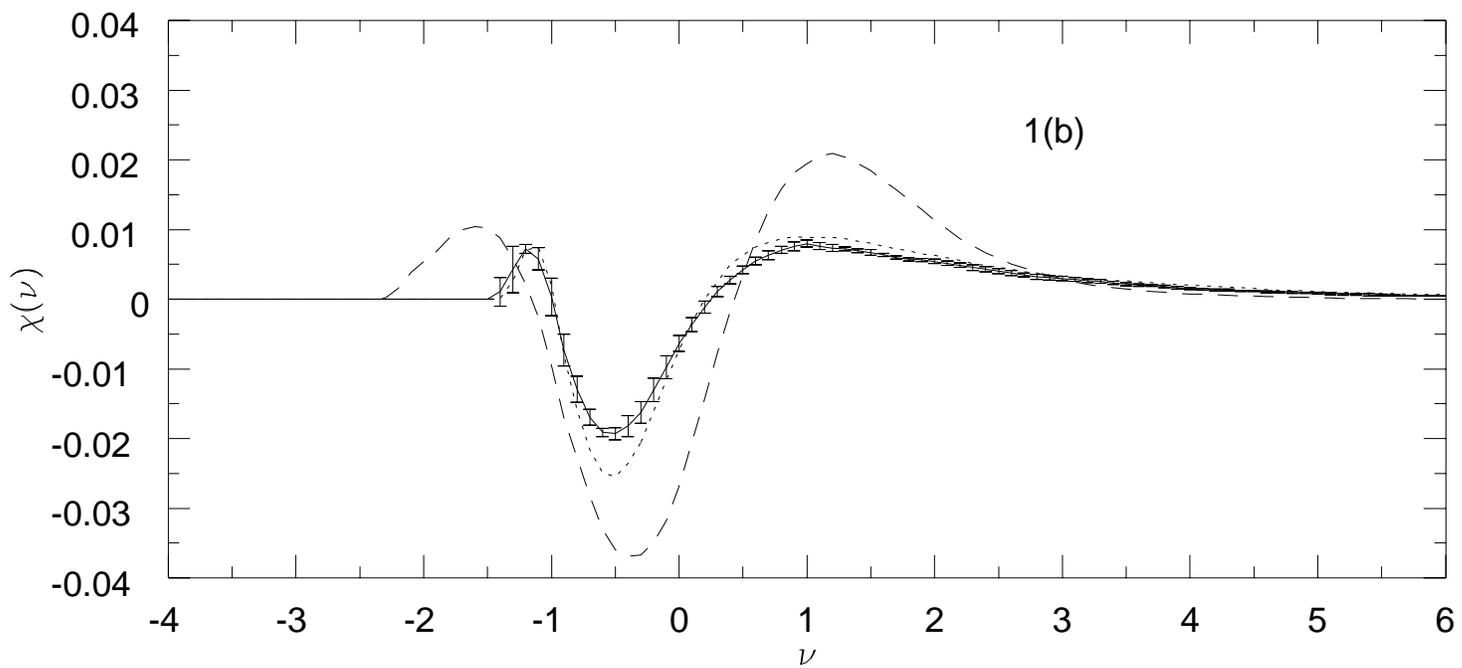
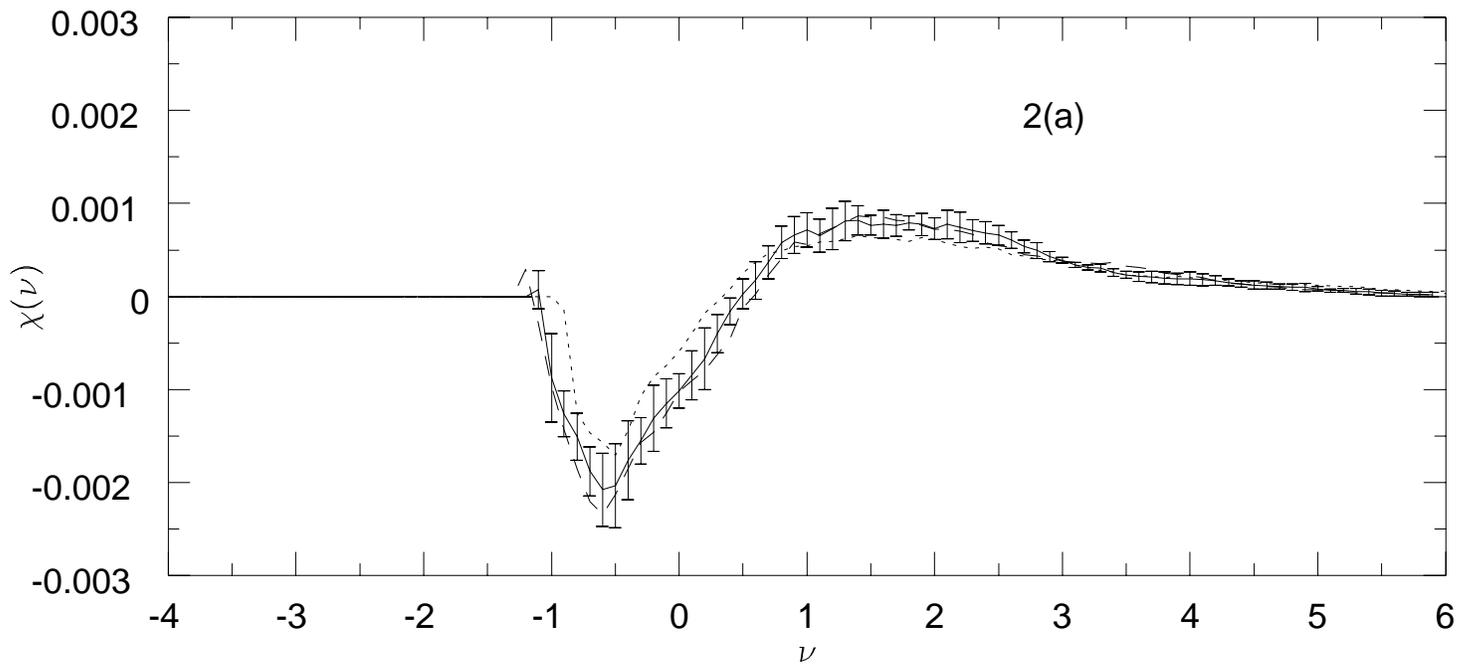
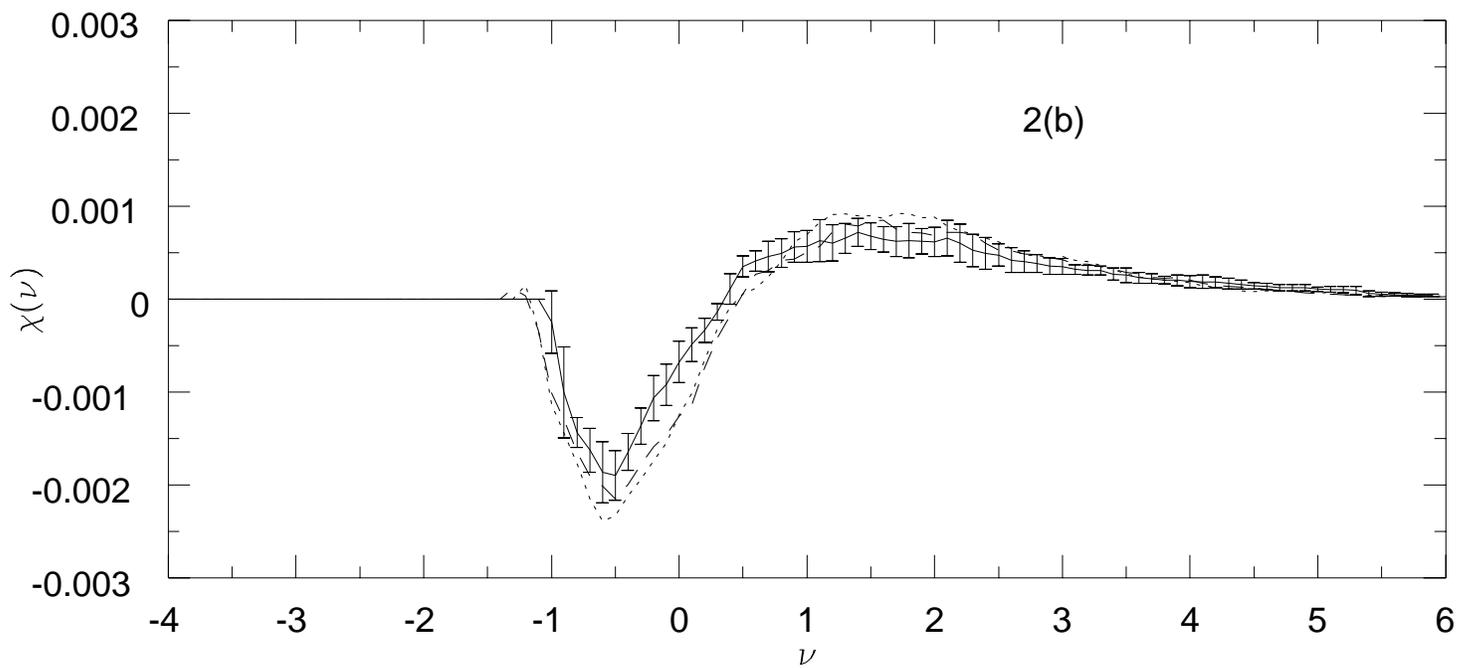
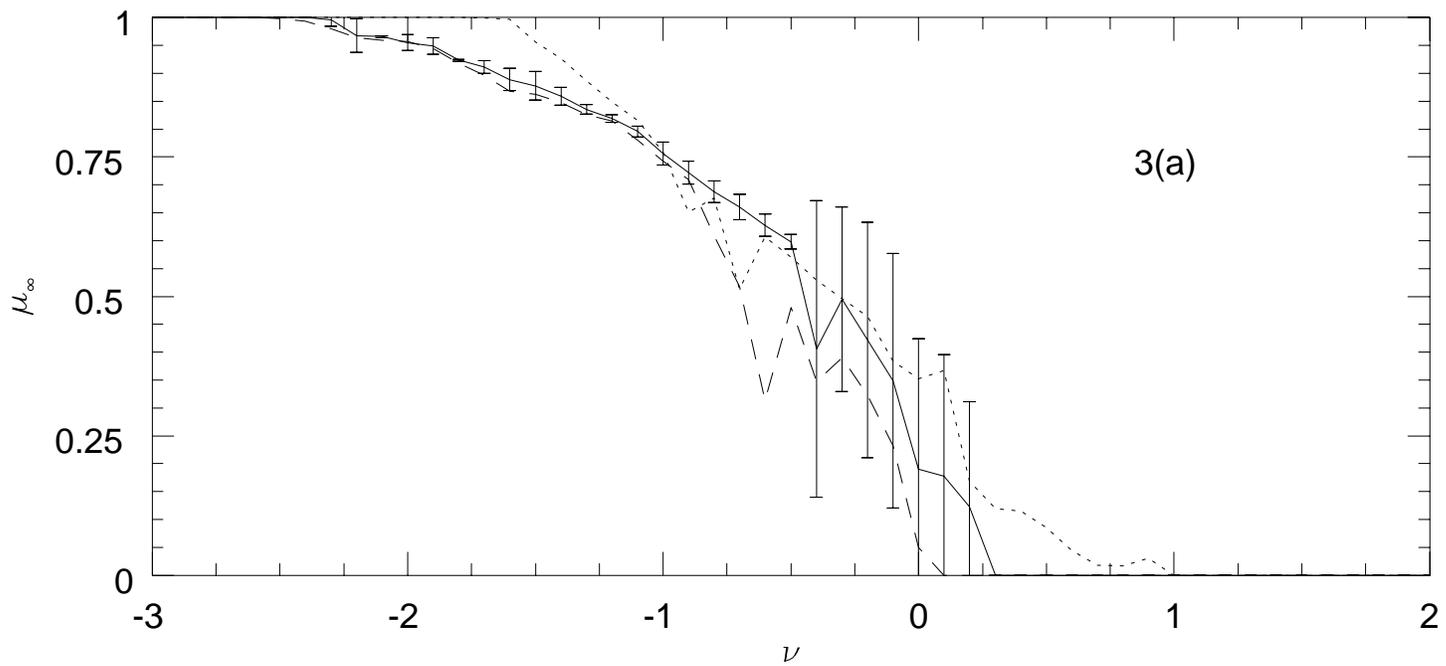
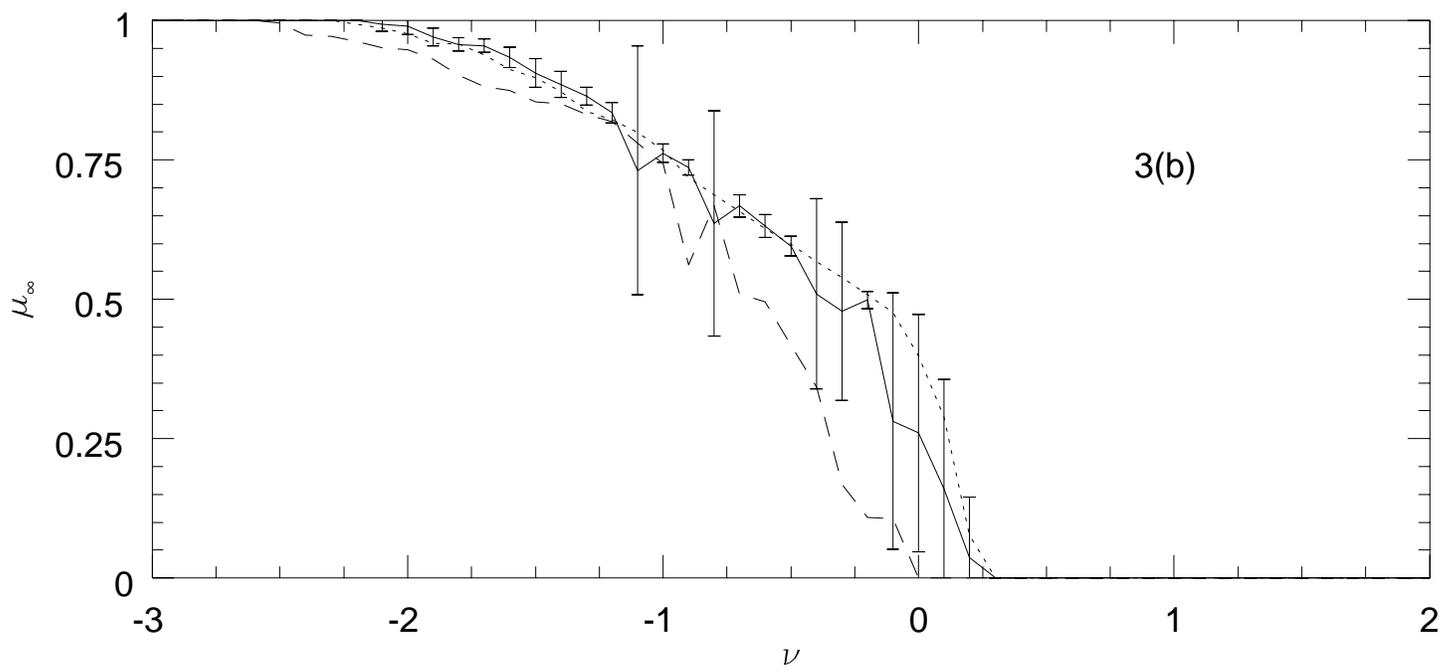
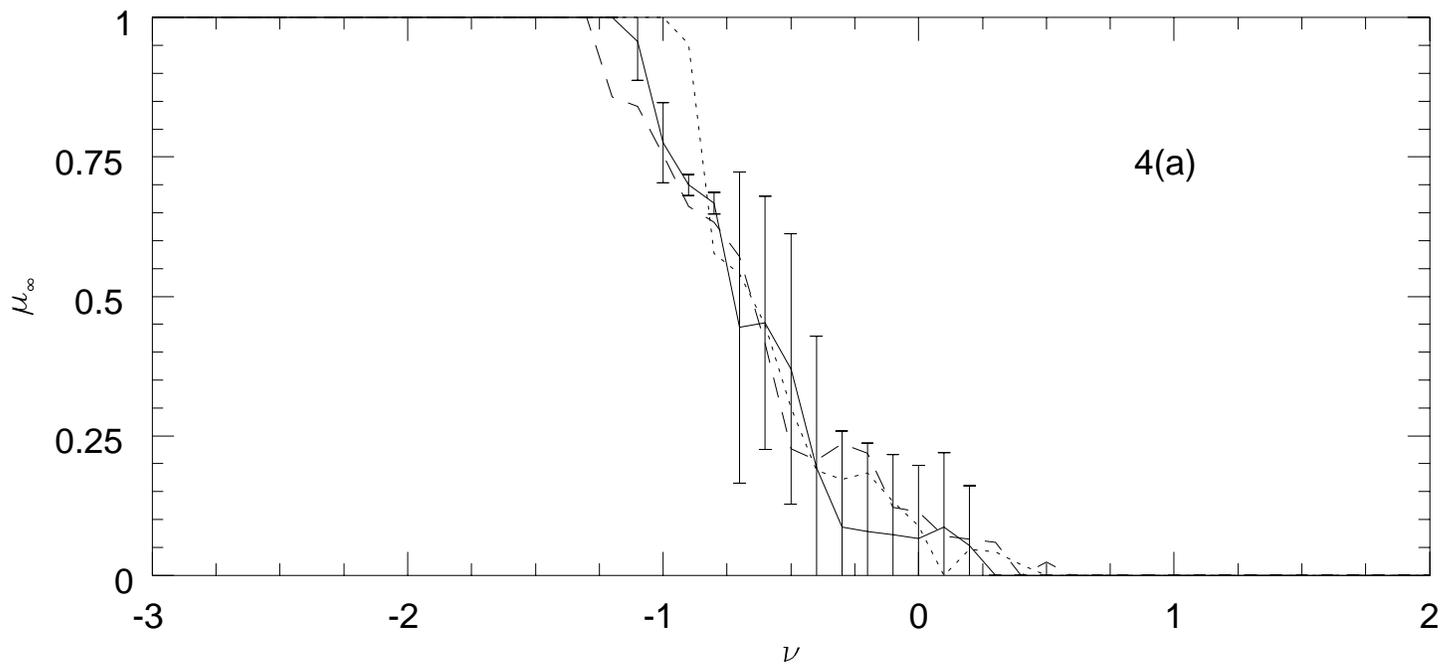
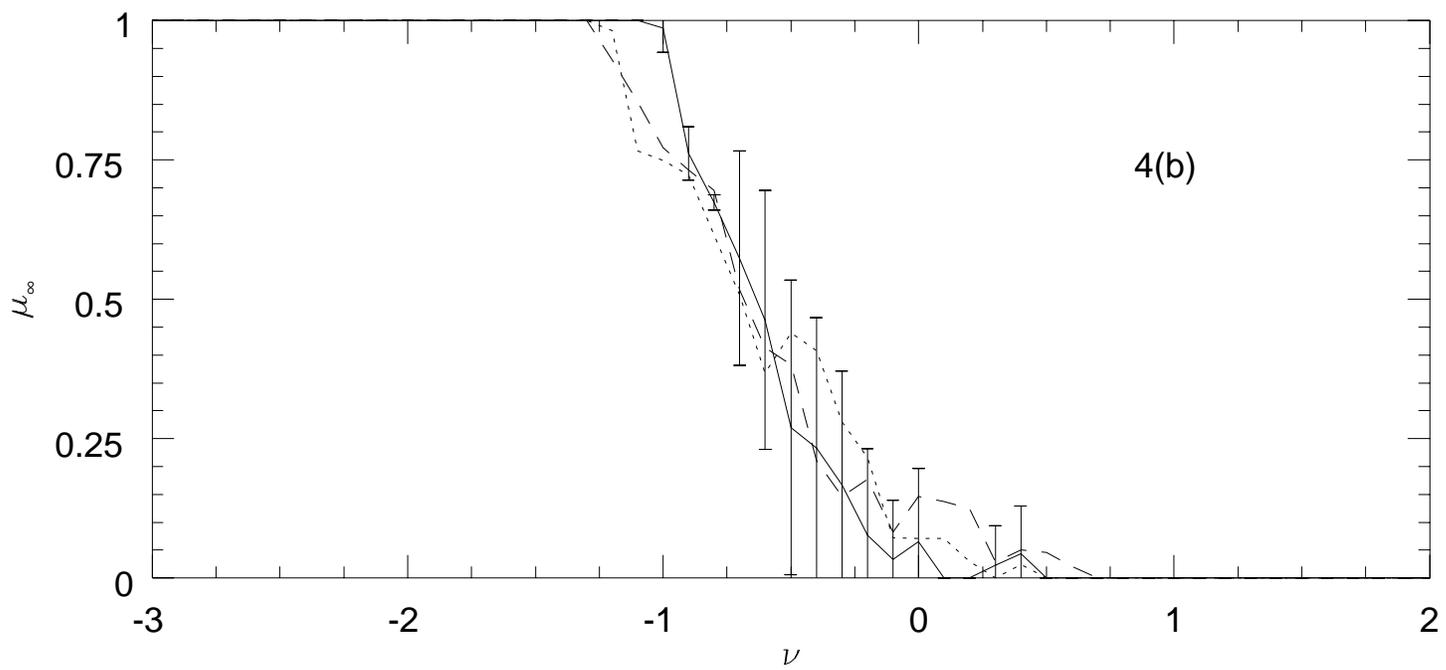
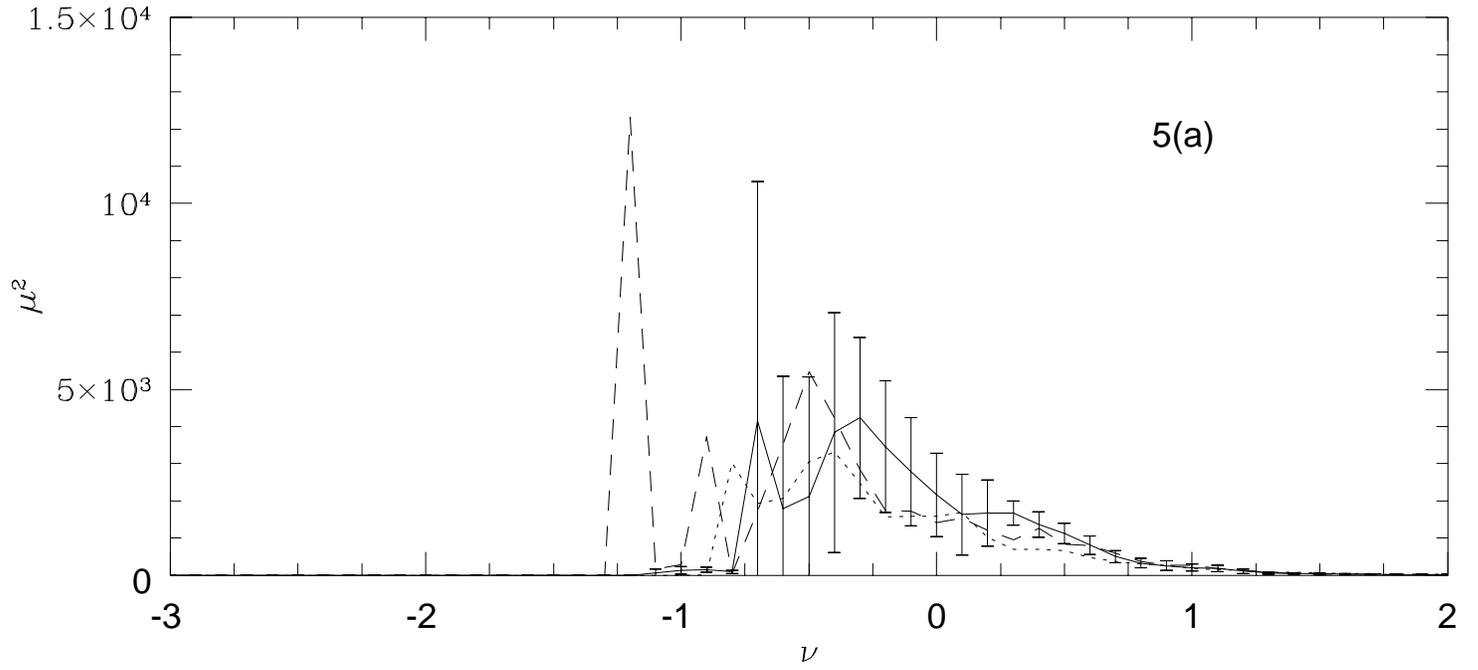
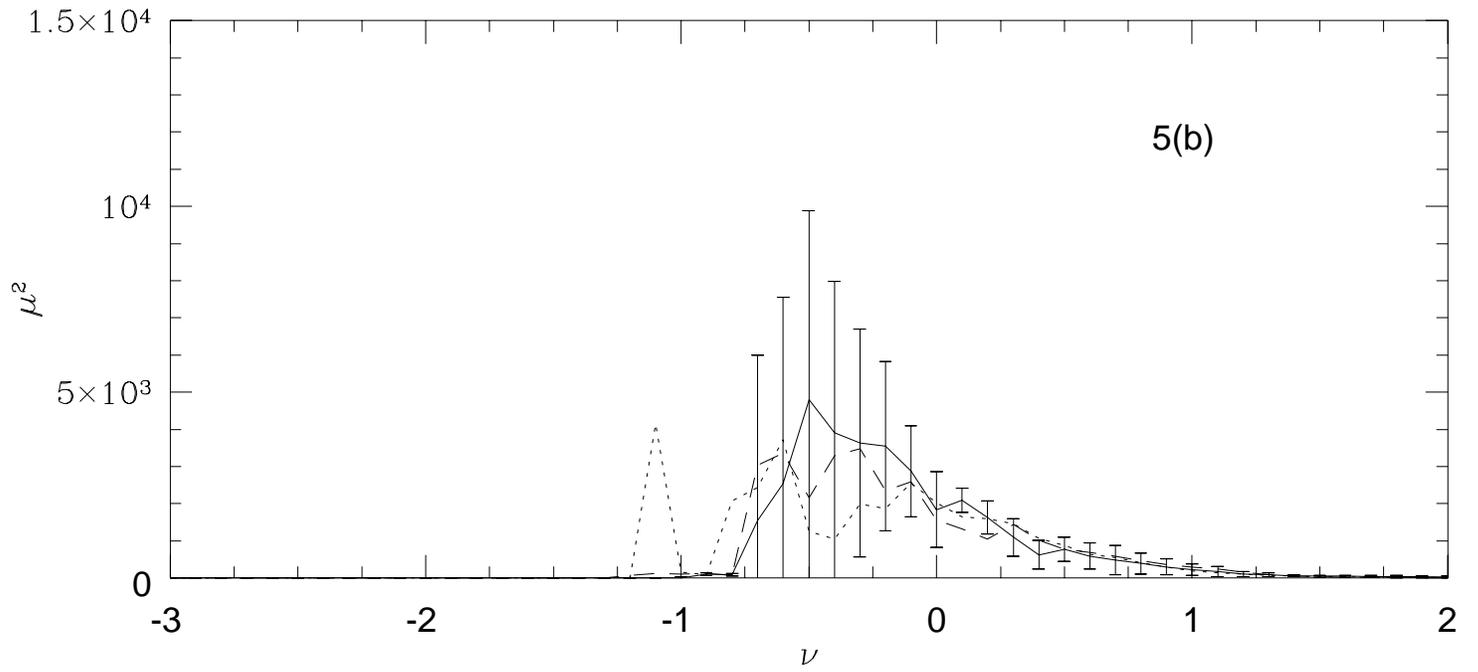
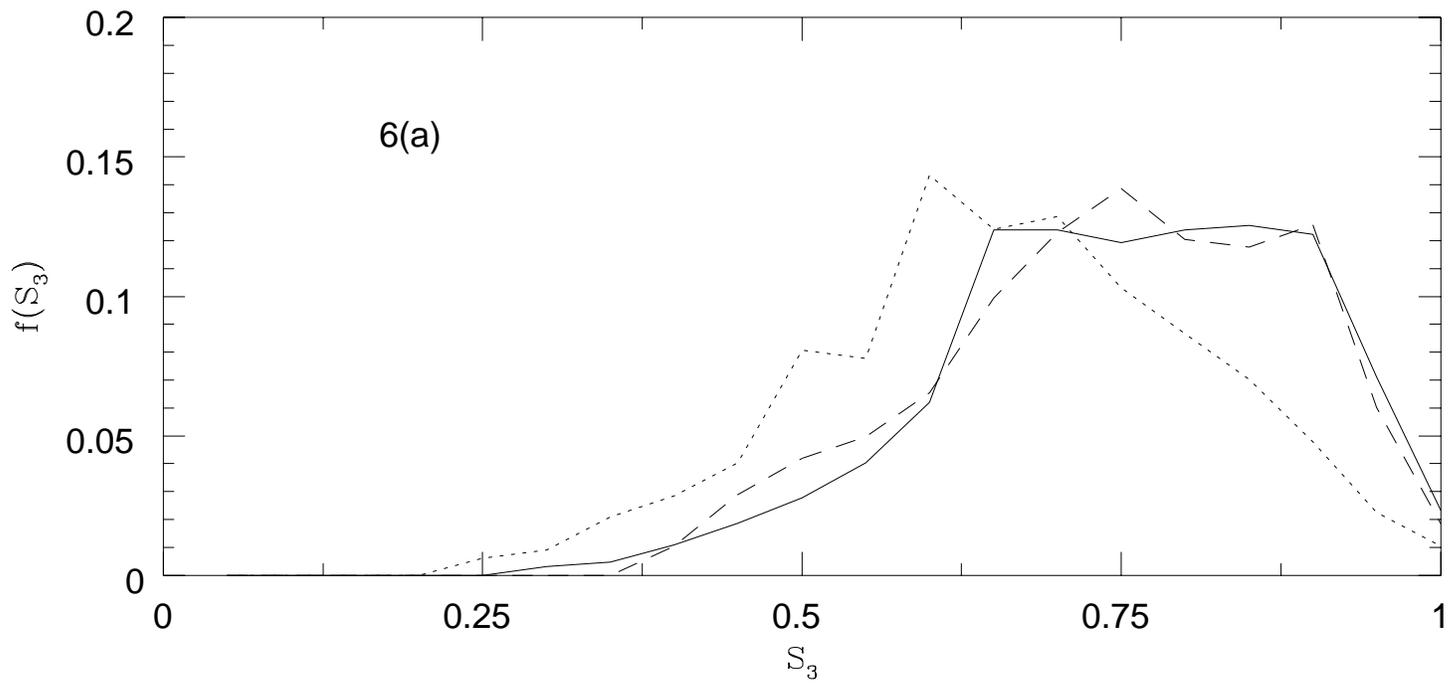
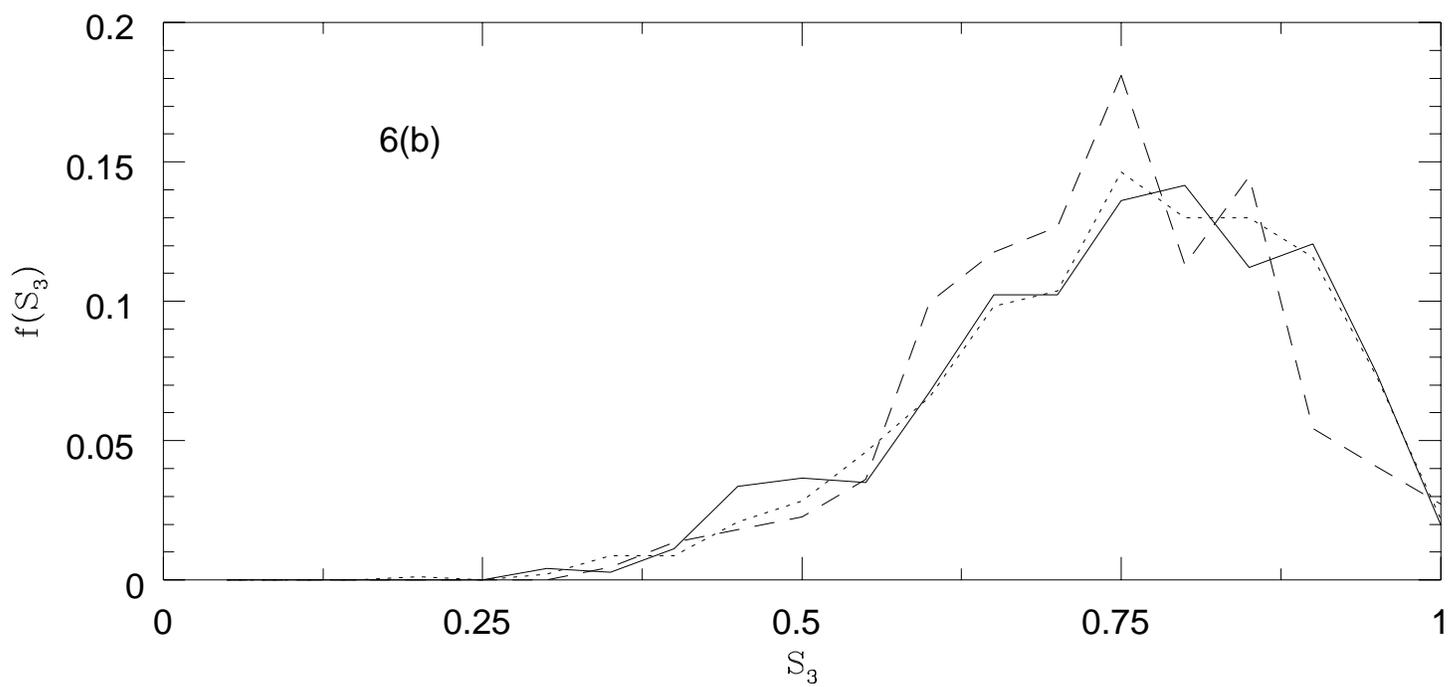
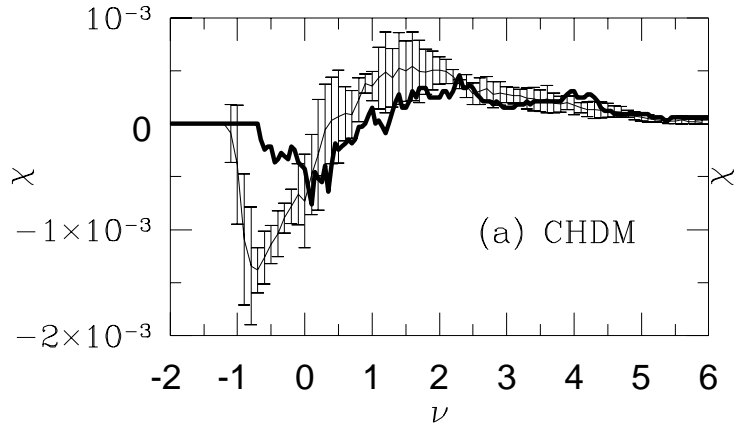
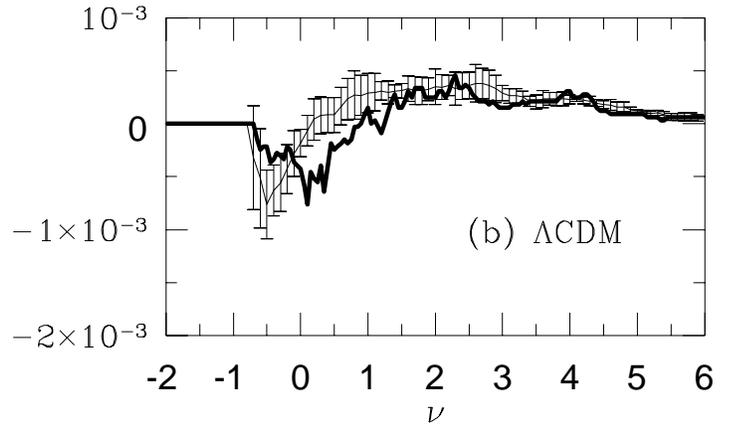
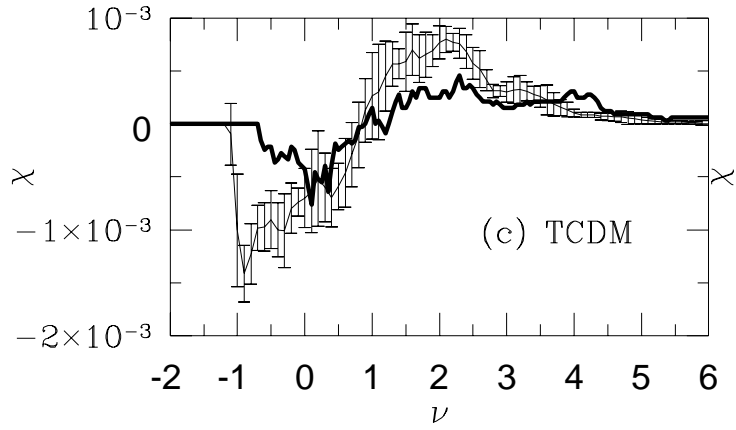
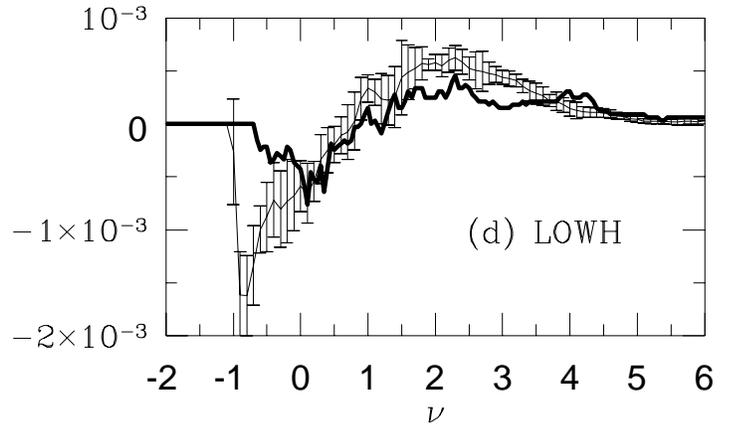
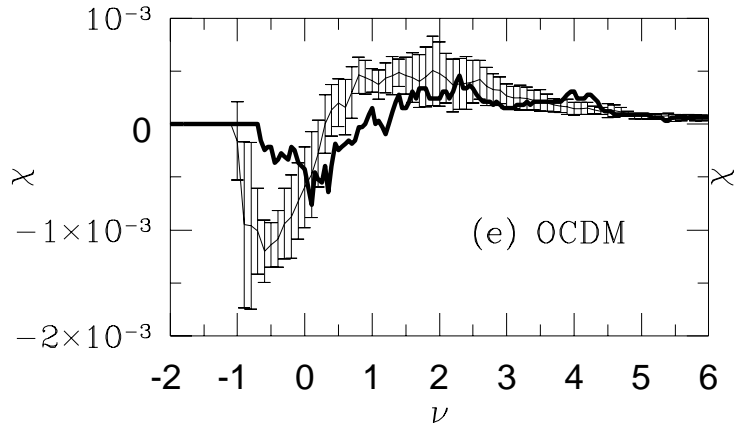
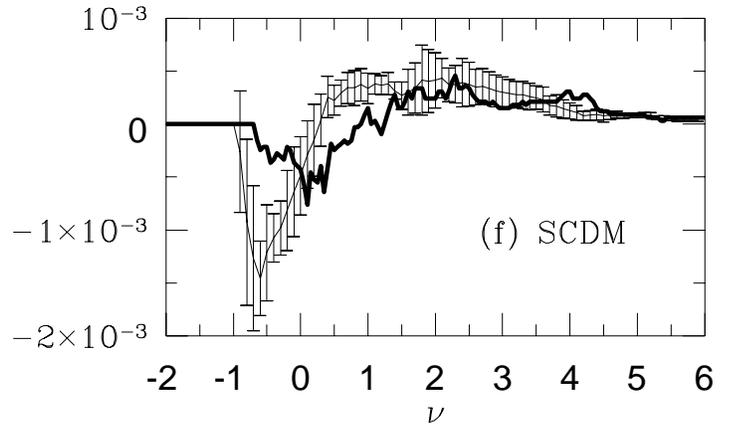
\end{document}